\documentclass[12pt]{iopart}
\usepackage{cite}
\usepackage{iopams}
\usepackage{multirow}
\usepackage{graphicx}
\usepackage{bm}
\usepackage{caption}
\usepackage{subcaption}
\usepackage[colorlinks, citecolor = blue, urlcolor = blue]{hyperref}

\expandafter\let\csname equation*\endcsname\relax
\expandafter\let\csname endequation*\endcsname\relax
\usepackage{amsmath}

\newcommand{\pto}[1]{\left( #1 \right)}
\newcommand{\pq}[1]{\left[ #1 \right]}

\newcommand{\bra}[1]{\left\langle #1 \right\vert}
\newcommand{\ket}[1]{\left\vert #1 \right\rangle}

\newcommand{\Braket}[2]{\left\langle #1 \bigg\vert #2 \right\rangle}

\begin{document}
\title{Quantum-limited estimation of the difference between photonic momenta via spatially resolved two-photon interference}

\author{Luca Maggio}
\address{School of Mathematics and Physics, University of Portsmouth, Portsmouth PO1 3HF, United Kingdom}
\address{Quantum Science and Technology Hub, University of Portsmouth, Portsmouth PO1 3FX, United Kingdom}
\author{Vincenzo Tamma}
\address{School of Mathematics and Physics, University of Portsmouth, Portsmouth PO1 3HF, United Kingdom}
\address{Quantum Science and Technology Hub, University of Portsmouth, Portsmouth PO1 3FX, United Kingdom}
\address{Institute of Cosmology and Gravitation, University of Portsmouth, Portsmouth PO1 3FX, United Kingdom}
\ead{vincenzo.tamma@port.ac.uk}

\vspace{10pt}
\begin{indented}
\item[]March 2026
\end{indented}

\begin{abstract}We present a quantum sensing protocol for three-dimensional estimation of the difference between the momenta of two photons based on spatially resolved interferometric sampling measurements. The protocol attains ultimate quantum precision in the simultaneous estimation of the components of the relative momentum for any values of the parameters already with $\sim 2000$ sampling measurements and a bias below $1\%$. These results identify 3D spatially resolved two-photon interference as an efficient tool for multi-parameter quantum sensing, with potential applications in high-precision 3D localization, refractometry, and near-field calibration in free-space quantum technologies.\end{abstract}

%
%
%
%
%

\section{Introduction}\label{sec:intro}
When two photons impinge in a balance beam splitter, interference can occur based on their identicality~\cite{PhysRevLett.59.2044,shih1988new}. In particular, if the photons are completely identical in all their degrees of freedom, they will always bunch on the same output port. This happens because, in case of fully overlapping photon wave packets, the events in which the two photons arrive in the two different output ports, the so-called coincidence events, interfere distructively. In general, it is possible to retrieve the overlap between the two-photon wave packets by counting the coincidence rate. If we encode in the overlap of the two photons specific parameters, this simple interferometric effect becomes the basis for a sensing protocol, which has already been used to estimate temporal delays~\cite{lyons2018attosecond, triggiani2023ultimate}, frequency shifts~\cite{PhysRevA.91.013830,Jin:15,Gianani_2018,fabre2021parameter,maggio2025quantum1}, spatial displacements~\cite{triggiani2024estimation,muratore2024superresolution,maggio2026ultimate}, or polarization parameters~\cite{harnchaiwat2020tracking,sgobba2023optimal,maggio2025multi}. The scalability, feasibility, and versatility of this protocol make it well suited for a wide range of applications, including quantum computing~\cite{kok2007linear,barz2012demonstration}, quantum key distribution~\cite{tang2014measurement,guan2015experimental}, quantum repeaters~\cite{sangouard2011quantum,hofmann2012heralded}, and quantum coherence tomography~\cite{teich2012variations}.

A central challenge in both fundamental research and technological applications is the simultaneous estimation of all the components of the momentum of photons emitted by individual single-photon sources, as it provides not only information about the photon’s energy but also about its direction of propagation. Several protocols have been developed to estimate the transverse momentum of single photons, typically based on detecting the photons in the far field, where transverse positions are directly mapped to transverse momentum~\cite{edgar2012imaging,freericks2023measure}. However, the achievement of the ultimate quantum sensitivity with these protocols has not been reached yet. Furthermore, the development of technologies enabling the simultaneous estimation of all momentum components concurrently while approaching the ultimate limits of quantum precision remains still terra incognita.

In this work, we introduce a sensing protocol for three-dimensional estimation of the difference between the momenta
of two photons based on interferometric sampling measurements. We demonstrate for the first time the ultimate quantum sensitivity in the simultaneous estimation of any values of the three components of the relative momenta, including angular differences. Applications include 3D imaging, localization and tracking of biological samples and high-precision refractometry~\cite{barer1957refractometry,wang2025ultrasensitive}. This two photon interference technique allows us also to probe photosensitive samples that could be damaged if exposed to a large number of photons~\cite{schermelleh2019super} and makes our protocol robust against phase fluctuations caused by external effects that can affect the precision in other protocols~\cite{PhysRevApplied.14.024028,PhysRevResearch.2.033191}. Furthermore, our technology relies in spatial-resolved measurements that are employed just after the two output ports of the  beam splitter, which makes our protocol compact and useful in free-space communication, where it is essential to calibrate the photonic directions in the near-field before free-space transmission begins~\cite{de2023satellite}. 

We show that our protocol can achieve ultimate quantum precision already with $\sim2000$ sampling measurements and independently of the value of the parameters to estimate. We also prove that measuring all the three components of the positions of the two photons in the near field at the output of the beam splitter increases the indistinguishability of the two photons and therefore enhance the sensitivity even when we are interested in the estimation of only one parameter associated with the photonic momenta.  

\section{Experimental setup}\label{sec:setup}
\begin{figure}
\centering
\includegraphics[width=100mm]{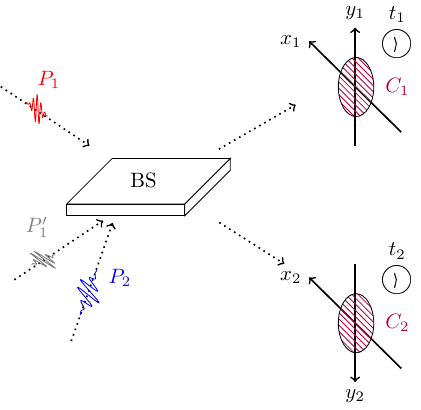}
\caption{Three-parameter quantum sensing scheme for the three-parameter estimation of the components of the relative momenta of two input photons. Two photons, $P_1$ and $P_2$, with different colors and approximately identical distribution in their transverse positions and emission times, interfere on a balanced beam splitter from two different angulations. In order to show the relative angulation of the two photons, we also plotted the symmetric image $P'_1$ of the first photon with respect to the plane of the beam splitter. The two-photon input state is represented by Eq.~\eqref{maineq:input}. After the beam splitter, their relative transverse position and temporal delay are measured in the near-field by the cameras $C_1$ and $C_2$.}\label{fig:setup}
\end{figure}

The aim of this section is to describe the sensing protocol represented in FIG.~\ref{fig:setup}. Two non-entangled photons are placed in the two different input ports of a balanced beam splitter. These two photons can differ both in their colors and in their 2D transverse momenta. They have identical Gaussian distributions in their transverse positions and emission times. Also, both photons are prepared so that they impinge on the same point of the beam splitter. Because of that, the probability amplitude of the $j$-th photon at coordinates $\vec{r}_j=(x_j,y_j,ct_j)$, where $t_j$ is the central time the photon $P_j$ impinges on the beam splitter, $c$ is the speed of light, and $(x_j,y_j)$ are the transverse positions of the j-th emitter, is 
\begin{eqnarray}
\eqalign{
     \psi_j(\vec{r}_j)&=\sqrt[4]{\frac{1}{(2\pi)^3\det(\Sigma_j)}}\exp\left[-\frac{(\vec{r}_j-\vec{\mu})^T\Sigma_j^{-1}(\vec{r}_j-\vec{\mu})}{4}\right]\\&\times\exp\left[-i\vec{r}^T_j\vec{k}_j\right],\label{maineq:amplt}
}
\end{eqnarray}
where, $\vec{k}_j=(-k_{x,j},-k_{y,j},\omega_j/c)$ defines the expected values of the frequency $\omega_j$ and the two transverse components of the momentum $k_{x,j},k_{y,j}$~\cite{Grynberg_Aspect_Fabre_2010}. For simplicity, we assume that the photonic transverse momenta components are much smaller than the longitudinal ones, namely $k_{z,j}$, so that $k_{z,j}\gg \sqrt{k_{x,j}^2+k_{y,j}^2}$, with $j=1,2$. Therefore, the paraxial approximation can be used to define the frequency of each photon, which reads $\omega_j/c\simeq k_{z,j}+(k_{x,j}^2+k_{y,j}^2)/2k_{z,j}$ for $j=1,2$. This applies to practical experimental scenarios where the numerical aperture is small and the beam waist of the photonic spatial distribution is smaller than the photonic wavelength~\cite{tragaardh2015simple,yew2013temporally,Nemoto1990NonparaxialGB}.
Given the states of the two photons
\begin{eqnarray}
    \eqalign{
        \ket{\Phi}&=\ket{\Phi_1}\otimes\ket{\Phi_2}\\&=\int d^3r_1 \psi_1(\vec{r}_1)a^\dagger_1(\vec{r}_1)\ket{0}\\&\otimes\int d^3r_2 \psi_2(\vec{r}_2)d^\dagger_2(\vec{r}_2)\ket{0},\label{maineq:input}
    }
\end{eqnarray}
where $a^\dagger_1(\vec{r}_1)$ is the creation operator for the photon in mode 1 and $d^\dagger_2(\vec{r}_2)$ is the creation operator for the photon in mode 2, the vector $\vec{\mu}$, which is the expected value of $\vec{r}_j$, for $j=1,2$, i.e.,
\begin{equation}
  \vec{\mu}=\langle\vec{r}_j\rangle=(\langle x_j\rangle,\langle y_j\rangle,c\langle t_j\rangle), 
\end{equation}
represents the average position and time of where both photons impinge on the beam splitter, and, without loss of generality, we define $\vec{\mu}=(0,0,0)$. The $j$-th three-dimensional covariance matrix of the spatial wave packet of the $j$-th photon is $\Sigma_j$, which reads
\begin{eqnarray}
    \eqalign{
        \Sigma_j&=\langle(\vec{r}_j-\mu)(\vec{r}_j-\mu)^T\rangle\\
        &=\begin{pmatrix}
            \Sigma_{x_j,x_j}&\Sigma_{x_j,y_j}&\Sigma_{x_j,ct_j}\\
            \Sigma_{x_j,y_j}&\Sigma_{y_j,y_j}&\Sigma_{y_j,ct_j}\\
            \Sigma_{x_j,ct_j}&\Sigma_{y_j,ct_j}&\Sigma_{ct_j,ct_j}
        \end{pmatrix}.\label{eq:covTammarequest}
    }
\end{eqnarray}
Several physical mechanisms can generate off-diagonal elements in the covariance matrix of a single-photon 3D Gaussian spatial distribution by coupling different spatial directions. In nonlinear sources like SPDC, phase-matching conditions related to the properties of the pump may link transverse and longitudinal momenta, producing correlations between axes~\cite{lee2016spatial}. Anisotropic pump profiles, such as elliptical or astigmatic beams, induce correlations between transverse directions, effectively rotating or tilting the photon’s spatial Gaussian relative to the lab axes~\cite{patil2023anisotropic}. Also, astigmatic or misaligned optical systems distort the spatial mode, rotating the principal axes and generating off-diagonal covariance terms~\cite{alda2003laser}. In Eq.~\eqref{eq:covTammarequest}, we consider these off-diagonal terms to take into consideration the most general setting, where these phenomena arise.

Because these effects influence the spatial structure of the photon, preparing photons with identical spatial wave-packets becomes a major experimental challenge. Even when using high-quality sources, the photons might not be fully identical in their positions and times. To formalize this, we assume that the covariance matrices of the two input photons, $\Sigma_1$ and $\Sigma_2$, are sufficiently similar and define their difference as $\delta\Sigma = \Sigma_2 - \Sigma_1$, with $\delta\Sigma \ll \Sigma_1, \Sigma_2$. Here, inequalities are understood in terms of the matrix norm~\cite{kato2013perturbation}. This allows us to treat the photons as nearly identical while accounting for residual discrepancies in their spatial correlations~\cite{pont2022quantifying,ding2025high}.   

Since the two photons have different transverse momenta $k_{x,j},k_{y,j}$, they will impinge at the beam splitter with a different angle, as represented in Fig.~\ref{fig:setup}. In order to fully define the input state, we define the two orthogonal bosonic operators $\hat{a}^\dagger_{i}\pto{\vec{r}}$ and $\hat{b}^\dagger_{i}\pto{\vec{r}}$, with $i=1,2$, obeying the commutation relations
\begin{eqnarray}
    \eqalign{
&\pq{\hat{a}_i\pto{\vec{r}_i},\hat{a}^\dagger_j\pto{\vec{r}_j}}=\delta_{i,j}\delta^3\pto{\vec{r}_i-\vec{r}_j},\\
&\pq{\hat{b}_i\pto{\vec{r}_i},\hat{b}^\dagger_j\pto{\vec{r}_j}}=\delta_{i,j}\delta^3\pto{\vec{r}_i-\vec{r}_j},\\
&\pq{\hat{b}_i\pto{\vec{r}_i},\hat{a}^\dagger_j\pto{\vec{r}_j}}=0.
    }
\end{eqnarray}
The operator $\hat{a}^\dagger_{1}\pto{\vec{r}}$ is associated with the photon in input channel 1 while, to take into account the differences in the states of the two photons in any degree of freedom apart from momentum and position, the mode operator of the photon in input channel 2 is defined as $\hat{d}^\dagger_{2}\pto{\vec{r_2}}=\sqrt{\nu}\hat{a}^\dagger_{2}\pto{\vec{r_2}}+\sqrt{1-\nu}\hat{b}^\dagger_{2}\pto{\vec{r_2}}$, where $\nu\in[0,1]$ defines the distinguishability of the two photons in these parameters. 

After impinging at the beam splitter, the two photons can exit the same output port or exit the two different output ports separately. We define these two events as bunching event and coincidence event, respectively. In order to "quantum erase" the distinguishability of the two photons in their momenta vectors, we introduce at the output of the beam splitter, in the near-field, two cameras, $C_1$ and $C_2$, resolving both the temporal delay of the two photons, namely $\Delta t$, and the difference in transverse position, defined by the two parameters $\Delta x,\Delta y$. Therefore, the cameras enable us to resolve the relative vector $\Delta \vec{r}=\vec{r}_2-\vec{r}_1=(\Delta x,\Delta y,c\Delta t)$.

\section{Output probabilities}\label{sec:output}
\begin{figure}
\centering
\includegraphics[width=100mm]{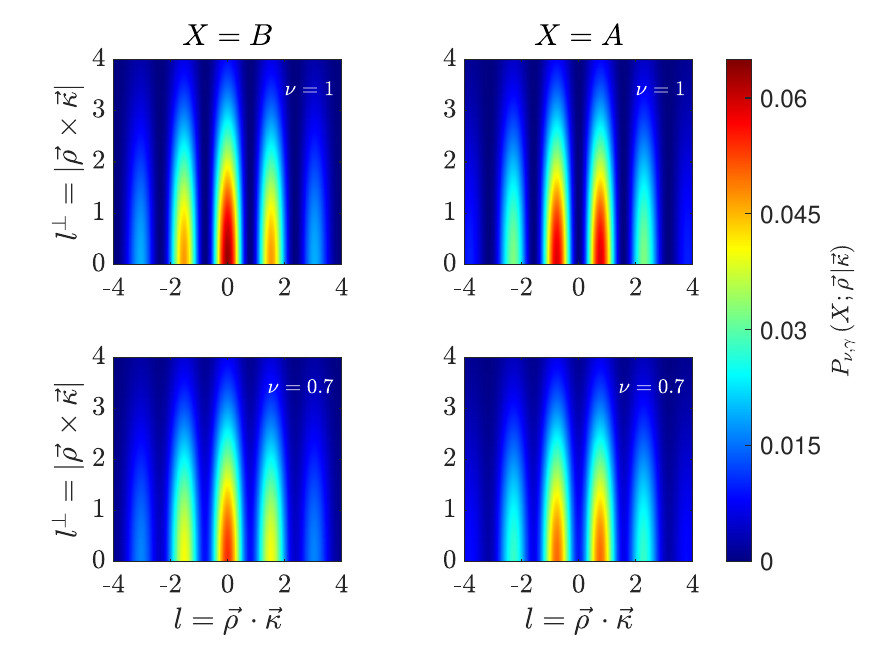}
\caption{Plot of the output probability distribution in Eq.~\eqref{maineq:prob}. For simplicity, the plot is made by imposing $\nu=1$ (top figures) and $\nu=0.7$ (bottom figures), $\gamma=1$ and $\vert\vec{\kappa}\vert=4$. As is it evident in Eq.~\eqref{maineq:prob}, the probability distribution presents a beating in the direction of $\vec{\kappa}$ that depends on $\vec{\kappa}\cdot\vec{ \rho}$ and on a Gaussian envelope that is function of $\vert\vec{ \rho}\vert^2=l^2+(l^{\perp})^2=(\vert\vec{\rho}\cdot\vec{ \kappa}\vert^2+\vert\vec{\rho}\times\vec{ \kappa}\vert^2)/\vert\vec{\kappa}\vert^2$, where the vectors $\vec{ \rho}$ and $\vec{\kappa}$ are defined in Eq.~\eqref{eq:reparameters}. Therefore, although the beatings oscillate only as a function of $l$, its modulation changes in both $l$ and $l^\perp$.
}\label{fig:probs}
\end{figure}

In this section, we will evaluate the output probabilities of the sensing protocol, as a function of the bunching and coincidence events, of the vector $\Delta \vec{r}$ and depending on the three-dimensional parameter $\Delta \vec{k}=\vec{k}_2-\vec{k}_1=(-\Delta k_x,-\Delta k_y,\Delta\omega/c)$, where $\Delta\omega$ is the difference in the photonic frequencies and $(\Delta k_x,\Delta k_y)$ are the components of the difference in the transverse momenta. The bunching and coincidence events are defined with the discrete variable $X$, where $X=A$ represents a coincidence event, while $X=B$ a bunching event. If the two photon state are prepared in such a way that the states appear almost identical in the 3D distributions in time and transverse position, i.e. $\delta\Sigma\ll\Sigma_1,\Sigma_2$, the corresponding probability distribution at the beam splitter output in the near field reads (see~\ref{app:evolution})
\begin{eqnarray}
    \eqalign{
          P_{\nu,\gamma}\left(X;\Delta \vec{r}|\Delta \vec{k}\right)&=\frac{\gamma^2}{\sqrt{(2\pi)^{3}\det(\Sigma)}}\exp\left[-\Delta\vec{r}^T\frac{\Sigma^{-1}}{2}\Delta\vec{r}\right]\\
        &\times\frac{1}{2}\Bigg\lbrace1+\alpha(X)\nu\cos\left(\Delta \vec{r}\cdot\Delta \vec{k}\right)\Bigg\rbrace,
    }\label{eq:P1}
\end{eqnarray}
where $\gamma\in[0,1]$ is the efficiency of each detector and $\alpha(A)=-1$ and $\alpha(B)=1$. Here, we define $\Sigma=\Sigma_1+\Sigma_2=2\Sigma_1+\delta\Sigma$, with eigenvalues $\sigma^2_k$, where $k=x,y,ct$. We define the relative canonical dimensionless variable $\vec{ \rho}$ and the parameter $\vec{\kappa}$
\begin{eqnarray}
    \eqalign{
        &\vec{ \rho}=\Sigma^{-1/2}\Delta\vec{r},\\
        &\vec{\kappa}=\Sigma^{1/2}\Delta\vec{k},
    }\label{eq:reparameters}
\end{eqnarray}
incorporating the positive semi-definite root $\Sigma^{-1/2}$ of the covariance matrix $\Sigma$ within the variable $\Delta\vec{r}$ and the parameter $\Delta\vec{k}$. In the case where $\Sigma$ is diagonal, we have
\begin{eqnarray}
    \eqalign{
        \vec{ \rho}&=(\Delta x/\sigma_x,\Delta y/\sigma_y,c\Delta t/\sigma_{ct}),\\
        \vec{\kappa}&=(-\sigma_x \Delta k_x,-\sigma_y \Delta k_y,\sigma_t\Delta\omega/c).
    }
\end{eqnarray}
Using $\Delta\vec{r}^T\Sigma^{-1}\Delta\vec{r}=\vert\vec{ \rho}\vert^2$ and $\Delta \vec{r}\cdot\Delta \vec{k}=\vec{ \rho}\cdot\vec{\kappa}$ in Eq.~\eqref{eq:P1}, it is possible to write the probability distribution as
\begin{eqnarray}
    \eqalign{
          P_{\nu,\gamma}\left(X;\vec{ \rho}| \vec{\kappa}\right)&=\frac{\gamma^2}{(2\pi)^{3/2}}\exp\left[-\frac{\vert\vec{ \rho}\vert^2}{2}\right]\zeta_{X;\nu}(\vec{ \rho}\cdot\vec{\kappa}),\label{maineq:prob}
    }
\end{eqnarray}
where
\begin{equation}
    \zeta_{X;\nu}(\vec{ \rho}\cdot\vec{\kappa})=\frac{1}{2}\Bigg\lbrace1+\alpha(X)\nu\cos\left(\vec{ \rho}\cdot\vec{\kappa}\right)\Bigg\rbrace.\label{maineq:quantumbeats}
\end{equation}
The output probability of Eq.~\eqref{maineq:prob} is plotted in Fig.~\ref{fig:probs} for two values of the distinguishability $\nu=1,0.7$. We see that along one direction of the space defined by $\vec{ \rho}$, it is possible to observe quantum beatings whose frequency is proportional to $|\vec{\kappa}|$ and direction defined by the direction of $\vec{\kappa}$. The intensity of the beating is modulated by a Gaussian envelope having mean equal to zero and unitary variance. The visibility of the beatings also depends on $\nu$ (maximum for $\nu=1$, minimum for $\nu=0$). The frequency of the beating and its direction define three parameters, that fully describe the vector $\vec{\kappa}$, and that can be estimated by sampling from the outcomes of the sensing protocol. In particular, we can define
\begin{equation}
    \vec{\kappa}=\vert\vec{\kappa}\vert\begin{pmatrix}
        \cos\theta\\\sin\theta\cos\phi\\ \sin\theta\sin\phi
    \end{pmatrix}.
\end{equation}
Recalling the definition in Eq.~\eqref{eq:reparameters} of the vector $\vec{\kappa}=((\Sigma^{1/2}\Delta\vec{k})_1,(\Sigma^{1/2}\Delta\vec{k})_2,(\Sigma^{1/2}\Delta\vec{k})_3)$, with $\Sigma$ defined in Eq.~\eqref{eq:covTammarequest} and $\vec{\Delta\vec{k}}=(-\Delta k_x,-\Delta k_y,\Delta\omega/c)$, we can define the parameters $(\vert\vec{\kappa}\vert,\theta,\phi)$ as 
\begin{eqnarray}
    \eqalign{
        \vert\vec{\kappa}\vert&=\sqrt{\Delta\vec{k}^T\Sigma\Delta\vec{k}}=\vert\Delta\vec{k}\vert\sqrt{\frac{\Delta\vec{k}^T\Sigma\Delta\vec{k}}{\vert\Delta\vec{k}\vert^2}},\\
         \theta&=
        \arccos\frac{(\Sigma^{1/2}\Delta\vec{k})_1}{\sqrt{\Delta\vec{k}^T\Sigma\Delta\vec{k}}},\qquad \theta\in[0,\pi],\\
        \phi&=\mathrm{sgn}((\Sigma^{1/2}\Delta\vec{k})_3)\arccos\frac{(\Sigma^{1/2}\Delta\vec{k})_2}{\sqrt{(\Sigma^{1/2}\Delta\vec{k})_2^2+(\Sigma^{1/2}\Delta\vec{k})_3^2}},\\&\qquad \phi\in[-\pi,\pi],
    }\label{eq:newparam}
\end{eqnarray}
which, in the case $\Sigma$ is diagonal, reduce to
\begin{eqnarray}
    \eqalign{
        \vert\vec{\kappa}\vert&=\sqrt{\sigma^2_x \Delta k^2_x+\sigma^2_y \Delta k^2_y+\frac{\sigma_t^2\Delta\omega^2}{c^2}},\\
         \theta&=
        \arccos\frac{-\sigma_x\Delta k_x}{\vert\vec{\kappa}\vert},\qquad \theta\in[0,\pi],\\
        \phi&=\mathrm{sgn}(\Delta \omega)\arccos\frac{-\sigma_y\Delta k_y}{\sqrt{\sigma_y^2\Delta k_y^2+\frac{\sigma_t^2\Delta\omega^2}{c^2}}},\, \phi\in[-\pi,\pi],
    }
\end{eqnarray}
respectively. These three parameters correspond to the modulus $\vert\vec{\kappa}\vert$ and direction ($\theta,\phi$) of the dimensionless vector $\vec{\kappa}$, which depends on the difference in transverse momenta and colors of the two photons. The modulus of $\vec{\kappa}$ depends linearly on the modulus of $\Delta \vec{k}$. However, it also depends on the Rayleigh quotient $\Delta\vec{k}^T\Sigma\Delta\vec{k}/\vert\Delta\vec{k}\vert^2$ of the matrix $\Sigma$ with respect to the vector $\Delta\vec{k}$, as we can see from Eq.~\eqref{eq:newparam}, due to the mixing and scaling of the components of $\Delta \vec{k}$ induced by the application of the matrix $\Sigma^{1/2}$~\cite{plemmons1988matrix}. This same mixing gives rise to a nontrivial mapping between the direction of $\vec{\kappa}$ and that of $\Delta \vec{k}$.
\section{Ultimate 3D quantum sensitivity}\label{sec:FI}
 In this section, we show what is the precision in the estimation of the parameters $(\vert\vec{\kappa}\vert,\theta,\phi)$ defining the vector $\vec{\kappa}$ by using this sensing protocol. It is important to notice that the parameters that we want to estimate are directly linked to the quantum beating. Hence, in order to maximize the precision in the estimation, the precision $\delta\vec{\rho}$ in measuring the 3D variable $\vec{\rho}$ in the near-field, $\delta\vec{ \rho}$, must be small enough to resolve the 3D spatial distribution and the quantum beatings, i.e.,
\begin{equation}
        |\delta \vec{ \rho}|\ll 1,\qquad\delta \vec{ \rho}\cdot\vec{\kappa}\ll 1.\label{eq:conditions}
\end{equation}
For the estimation of the three parameters $(\vert\vec{\kappa}\vert,\theta,\phi)$ we consider a set of likelihood estimators $(\tilde{\vert\vec{\kappa}\vert}, \tilde{\theta},\tilde{\phi})$. A set of unbiased estimators is affected by an error defined by the covariance matrix $\mathrm{Cov}[\tilde{\vert\vec{\kappa}\vert}, \tilde{\theta},\tilde{\phi}]$. This matrix is bounded from below by the Cram\'{e}r-Rao bound, which represent the maximum precision achievable by the sensing protocol. In particular, this bound is inversely proportional to the number of sampling measurement, namely $N$, and directly proportional to the inverse of the Fisher information matrix, that is $F^{-1}_\nu\pto{\vert\vec{\kappa}\vert,\theta,\phi}$~\cite{cramer1999mathematical,rohatgi2015introduction}. Asymptotically, i.e., in the limit of large $N$, the likelihood estimators are unbiased and the Cram\'{e}r-Rao bound is always saturable. However, while the Fisher information matrix represent the maximum information achievable from the sensing protocol, it is not necessarily the maximum information achievable from the probe. Indeed, it is possible to set an upper bound for the Fisher information matrix, which is the quantum Fisher information matrix, that is $Q^{-1}\pto{\vert\vec{\kappa}\vert,\theta,\phi}$~\cite{helstrom1969quantum,holevo2011probabilistic}. Altogether, these inequalities can expressed as follows
\begin{equation}
\mathrm{Cov}[\tilde{\vert\vec{\kappa}\vert}, \tilde{\theta},\tilde{\phi}]\geq \frac{F^{-1}_\nu\pto{\vert\vec{\kappa}\vert,\theta,\phi}}{N}\geq \frac{Q^{-1}\pto{\vert\vec{\kappa}\vert,\theta,\phi}}{N}. \label{maineq:bounds}
\end{equation}
If the expected values of the commutators of the symmetric logarithmic derivatives related to the parameters we want to estimate are equal to zero, then the maximum information achievable from the probe through the proposed quantum sensing protocol is equal to the Quantum Fisher information matrix~\cite{Liu_2020}. We define the symmetric logarithmic derivatives related to the parameters $\vert\vec{\kappa}\vert,\theta,\phi$ as $\hat{L}_{\star}$,where $\star=\vert\vec{\kappa}\vert,\theta,\phi$, by the formula
\begin{eqnarray}
    \eqalign{
        \frac{\partial \ket{\Phi}\bra{\Phi}}{\partial \star}&=\frac{\ket{\Phi}\bra{\Phi}\hat{L}_{\star}+\hat{L}_{\star}\ket{\Phi}\bra{\Phi}}{2}.
    }
\end{eqnarray}
The input state $\ket{\Phi}$ in Eq.~\eqref{maineq:input} is a pure state. Therefore, we can write
\begin{eqnarray}
\eqalign{
  \frac{\partial \ket{\Phi}\bra{\Phi}}{\partial \star}&=\frac{\partial (\ket{\Phi}\bra{\Phi})^2}{\partial \star}\\&=  \ket{\Phi}\bra{\Phi}\frac{\partial \ket{\Phi}\bra{\Phi}}{\partial \star}+\frac{\partial \ket{\Phi}\bra{\Phi}}{\partial \star}\ket{\Phi}\bra{\Phi},
  }
\end{eqnarray}
which allows us to write the symmetric logarithmic derivatives just by comparison with the previous set of equations, obtaining
\begin{eqnarray}
    \eqalign{
        \hat{L}_{\star}&=2\frac{\partial \ket{\Phi}\bra{\Phi}}{\partial \star},
    }
\end{eqnarray}
or, more explicitly,
\begin{eqnarray}
    \eqalign{
        &\hat{L}_{\star}=2\left(\ket{\frac{\partial\Phi}{\partial \star}}\bra{\Phi}+\ket{\Phi}\bra{\frac{\partial\Phi}{\partial \star}}\right),\label{app:defSLD}
    }
\end{eqnarray}
where
\begin{equation}
    \ket{\frac{\partial\Phi}{\partial\star}}=\frac{\partial}{\partial\star}\ket{\Phi}\qquad,\qquad \star=\vert\vec{\kappa}\vert,\theta,\phi.
\end{equation}
From Eq.~\eqref{app:defSLD}, we can prove that (see~\ref{SLD})
\begin{eqnarray}
    \eqalign{
        &\mathrm{Tr\left[\ket{\Phi}\bra{\Phi}\left[\hat{L}_{\vert\vec{\kappa}\vert},\hat{L}_\theta\right]\right]}=0,\\
        &\mathrm{Tr\left[\ket{\Phi}\bra{\Phi}\left[\hat{L}_{\theta},\hat{L}_\phi\right]\right]}=0,\\
        &\mathrm{Tr\left[\ket{\Phi}\bra{\Phi}\left[\hat{L}_{\phi},\hat{L}_{\vert\vec{\kappa}\vert}\right]\right]}=0,\\ \label{eq:SLD}
    }
\end{eqnarray}
implying that the Cram\'{e}r-Rao bound is saturable. 
 In particular, the quantum Fisher information matrix is (see~\ref{app:QFI})
\begin{eqnarray}
 \eqalign{
Q(\vert\vec{\kappa}\vert,\theta,\phi)&=\mathrm{Diag}[Q(\vert\vec{\kappa}\vert),Q(\theta),Q(\phi)]\\&=\begin{pmatrix}
        \left\vert\frac{\partial \vec{\kappa}}{\partial \vert\vec{\kappa}\vert}\right\vert^2&0&0\\0&\left\vert\frac{\partial \vec{\kappa}}{\partial \theta}\right\vert^2&0\\0&0&\left\vert\frac{\partial \vec{\kappa}}{\partial \phi}\right\vert^2
    \end{pmatrix}\\&=\begin{pmatrix}
        1&0&0\\0&\vert\vec{\kappa}\vert^2&0\\0&0&\vert\vec{\kappa}\vert^2\sin^2\theta
    \end{pmatrix}.  \label{eq:Q}   
 }   
\end{eqnarray}
The sensitivity to variations in each parameter $\vert\vec{\kappa}\vert,\theta,\phi$ is determined by the diagonal elements of the quantum Fisher information matrix in Eq.~\eqref{eq:Q}, which quantify the infinitesimal changes of the vector $\vec{\kappa}$ with respect to $\vert\vec{\kappa}\vert,\theta,\phi$, respectively. Eq.~\eqref{eq:Q} also admits a clear geometrical interpretation. For instance, the sensitivity to the parameter $\phi$, given by the diagonal term $\left\vert\frac{\partial \vec{\kappa}}{\partial \phi}\right\vert^2=\vert\vec{\kappa}\vert^2\sin^2\theta$ is maximized (minimized) at $\theta=\pi/2$ ($\theta=0,\pi$) and increases with $\vert\vec{\kappa}\vert^2$. Similarly, the sensitivity to $\theta$, $\left\vert\frac{\partial \vec{\kappa}}{\partial \theta}\right\vert^2=\vert\vec{\kappa}\vert^2$, grows with $\vert\vec{\kappa}\vert^2$ and is independent of both $\theta$ and $\phi$. In contrast, the sensitivity to $\vert\vec{\kappa}\vert$, remains constant regardless of the values of the other parameters.
We can obtain the Fisher information matrix from Eq.~\eqref{maineq:prob}. In particular, in the case of total indistinguishability  $\nu=1$, we obtain 
\begin{eqnarray}
F_{\nu=1}\pto{\vert\vec{\kappa}\vert,\theta,\phi}=\gamma^2Q\pto{\vert\vec{\kappa}\vert,\theta,\phi},\label{eq:finu1}
\end{eqnarray}
proving that, with perfectly efficient detectors indistinguishability of the photons in non-spatial and non-temporal parameters, we can achieve the maximum precision for the estimation of the parameters $(\vert\vec{\kappa}\vert,\theta,\phi)$.
\section{3D estimation in the case of partially
distinguishable photons at the detectors}
\begin{figure}
\centering
\includegraphics[width=100mm]{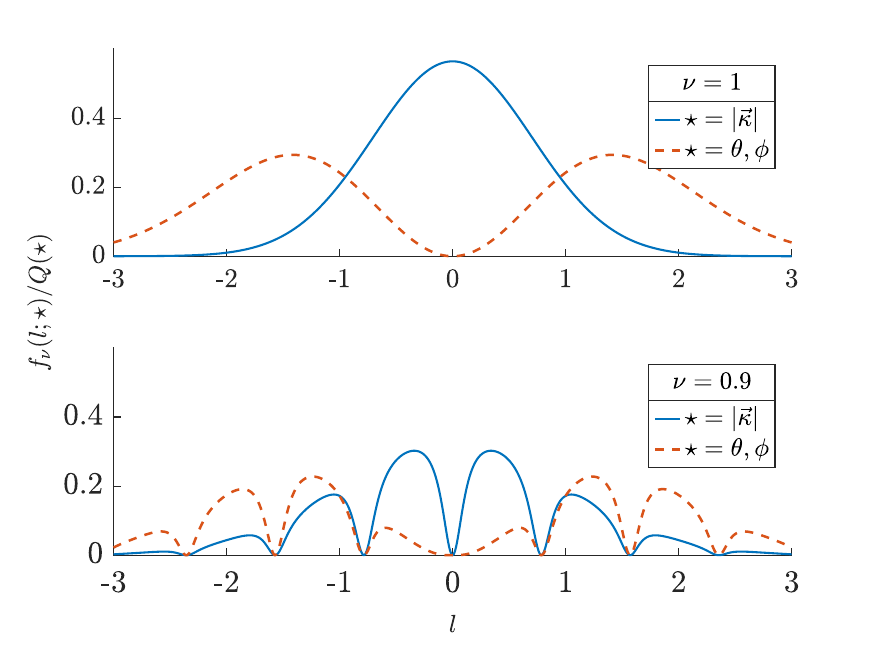}
\caption{Contributions of the Fisher information matrix density $f_{\nu}\pto{ l;\vert\vec{\kappa}\vert,\theta,\phi}$ in Eq~\ref{eq:densityfi} (normalized by the quantum Fisher information matrix $Q(\vert\vec{\kappa}\vert,\theta,\phi)$ in Eq.~\eqref{eq:Q}) as a function of $ l=\vec{ \rho}\cdot\vec{\kappa}/\vert\vec{\kappa}\vert$ for different values of the distinguishability parameter $\nu=1,0.9$ for photons with Gaussian spatial distributions $\vert\psi_i(\vec{r})\vert^2$, $i=1,2$, with $\vec{r}=(x,y,ct)$, where $t$ is the detection time, $c$ is the speed of light, and $(x,y)$ is the transverse position resolved by the cameras. For simplicity, we impose $\vert\vec{\kappa}\vert=4$.}\label{fig:fi}
\end{figure}
In this section, we consider the case of partially distinguishable photons, i.e., $\nu\neq 1$. In this case, the sensing protocol can still estimate the parameters $(\vert\vec{\kappa}\vert,\theta,\phi)$, however, the Cram\'{e}r-Rao bound is not saturated. This happens because, for partially distinguishable photons, the beating pattern is less pronounced. As a result, key features such as its direction and frequency—which directly encode the parameters we aim to estimate—are harder to infer by sampling from the probability distribution than in the fully indistinguishable case. The Fisher information matrix can be written as an integral with respect to the argument of the beating, i.e., $\vec{ \rho}\cdot\vec{\kappa}$. By renaming $\vec{ \rho}\cdot\vec{\kappa}/\vert\vec{\kappa}\vert= l$, we have (see~\ref{app:FI})
\begin{eqnarray}
F_{\nu}(\vert\vec{\kappa}\vert,\theta,\phi)&=\gamma^2\int d l f_\nu( l;\vert\vec{\kappa}\vert,\theta,\phi),\label{eq:Fi}
\end{eqnarray}
with the density of the Fisher information matrix 
\begin{eqnarray}
\eqalign{
    f_\nu( l;\vert\vec{\kappa}\vert,\theta,\phi)&=\mathrm{Diag}[f_\nu( l;\vert\vec{\kappa}\vert),f_\nu( l;\theta),f_\nu( l;\phi)]\\&= \frac{\mathrm{e}^{- l^2/2}\beta_\nu(\vert\vec{\kappa}\vert l)}{(2\pi)^{1/2}}\begin{pmatrix}
         l^2&0&0\\
        0&\vert\vec{\kappa}\vert^2&0\\
        0&0&\vert\vec{\kappa}\vert^2\sin^2\theta
    \end{pmatrix},\label{eq:densityfi}
}
\end{eqnarray}
which terms describe the metrological information for the parameters $\vert\vec{\kappa}\vert,\theta,\phi$ associated with each possible detected value of $l$  as plotted in FIG~\ref{fig:fi}.

Here, the function
\begin{eqnarray}
\eqalign{
    \beta_\nu(\vert\vec{\kappa}\vert l)&=\sum_{X=A,B}\frac{1}{\zeta_{X;\nu}(\vert\vec{\kappa}\vert l)}\left(\frac{\partial\zeta_{X;\nu}(\eta)}{\partial \eta}\right)_{\eta=\vert\vec{\kappa}\vert l}^2\\&=\frac{\nu^2\sin^2(\vert\vec{\kappa}\vert l)}{1-\nu^2\cos^2(\vert\vec{\kappa}\vert l)}
    }\label{eq:beta}
\end{eqnarray}
describes the contributions to the Fisher information from the quantum beats emerging in the expression of $\zeta_{X;\nu}(\vert\vec{\kappa}\vert l)$ in Eq.~\eqref{maineq:quantumbeats}. 

In Eq.~\eqref{eq:densityfi}, we can see that, similarly to Eq.~\eqref{eq:Q}, the sensitivity achievable in $\vert\vec{\kappa}\vert$ and $\theta$ does not depend on $\theta$ and $\phi$, while the sensitivity achievable in $\phi$ does not depend on $\phi$. However, each term of the Fisher information matrix depends on $\vert\vec{\kappa}\vert$, since the frequency of the beating varies as a function of this parameter. 

The function $\beta(\vert\vec{\kappa}\vert l)$ is the global degree of sensitivity at small variation of the value $\vert\vec{\kappa}\vert l$. If there photons are fully distinguishable at the detectors, i.e., if $\nu=0$, there is no beating, and no parameters can be estimated. In this case, $\beta(\vert\vec{\kappa}\vert l)=0$. Instead, if the photons are fully indistinguishable at the detectors, i.e., if $\nu=1$, the precision in the estimation of all parameters is maximum. Under this condition, $\beta(\vert\vec{\kappa}\vert l)=1$ and Eq.~\eqref{eq:Fi} reduces to Eq.~\eqref{eq:finu1}.

For values of $\nu\neq0,1$, the function $\beta(\vert\vec{\kappa}\vert l)$ oscillates with half the period and in the same direction of the beating of the probabilities in Eq.~\eqref{maineq:prob}. In fact, we have $(\partial \zeta_{X;\nu}(\eta)/\partial \eta)_{\eta=\vert\vec{\kappa}\vert l}^2=\nu^2\sin^2(\vert\vec{\kappa}\vert l)/4=\nu^2(1-\cos(2\vert\vec{\kappa}\vert l))/8$. In this case, we have $\beta(\vert\vec{\kappa}\vert l)=0$ for values of $\vert\vec{\kappa}\vert l$ that are multiples of $\pi$. These are the stationary points of the quantum beating in Eq.~\eqref{maineq:quantumbeats}, in which a small change in the value of $\vert\vec{\kappa}\vert$ does not significantly change the value of the probability distribution, with a consequent loss in sensitivity. As expected, measurements at these values of $ l$ do not provide any information in the estimation.

\begin{figure*}
\begin{minipage}[c]{0.48\linewidth}
\includegraphics[width=\linewidth]{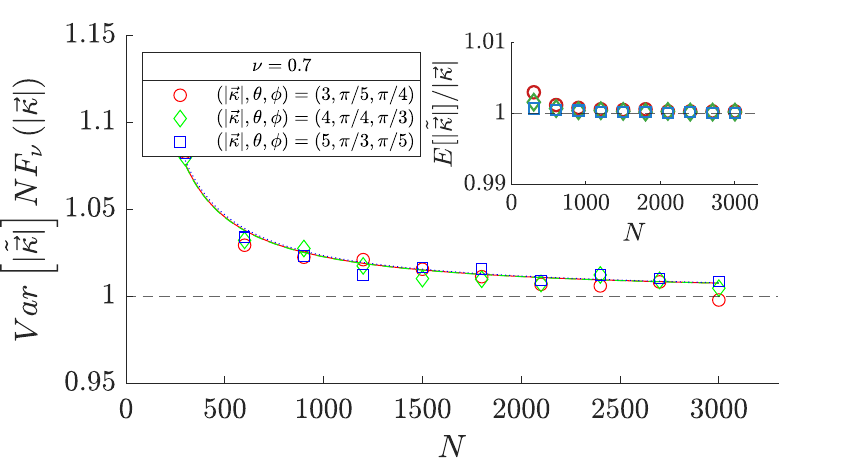}
\includegraphics[width=\linewidth]{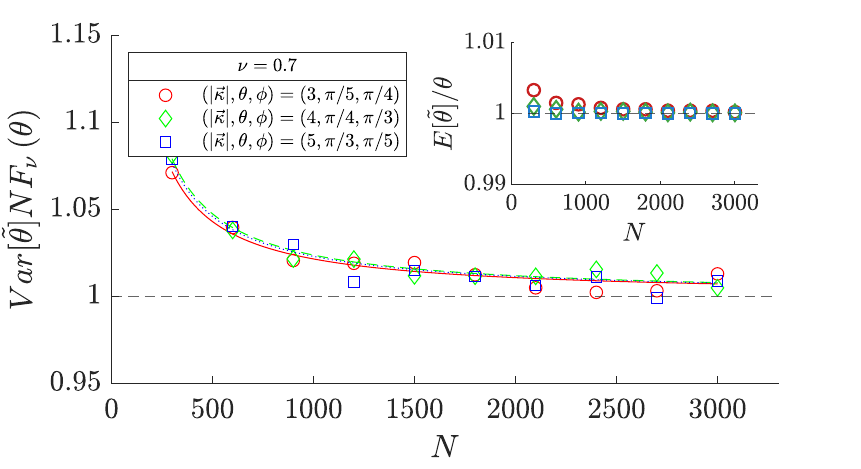}
\includegraphics[width=\linewidth]{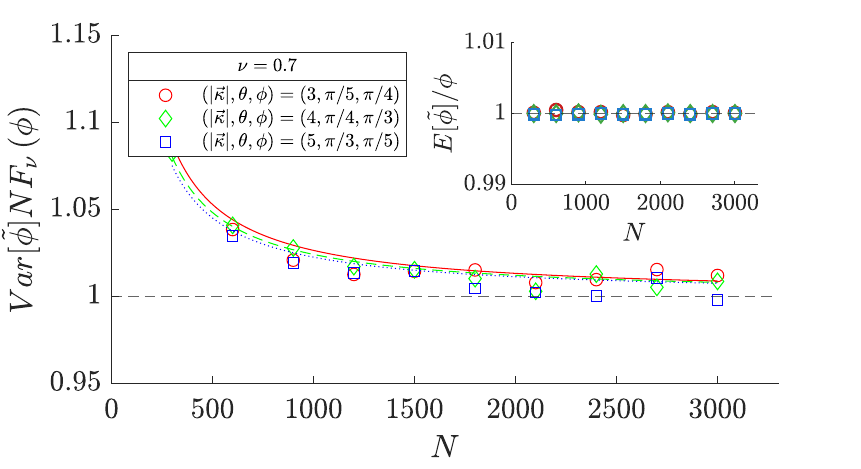}
\end{minipage}
\hfill
\begin{minipage}[c]{0.48\linewidth}

\includegraphics[width=\linewidth]{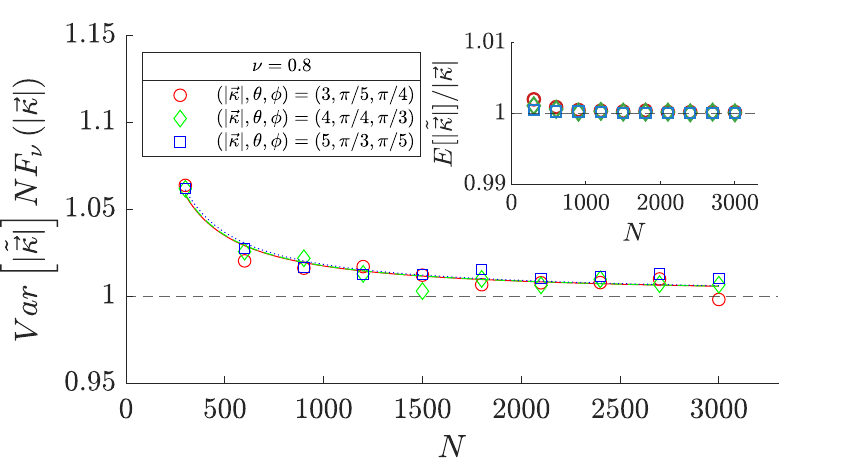}
\includegraphics[width=\linewidth]{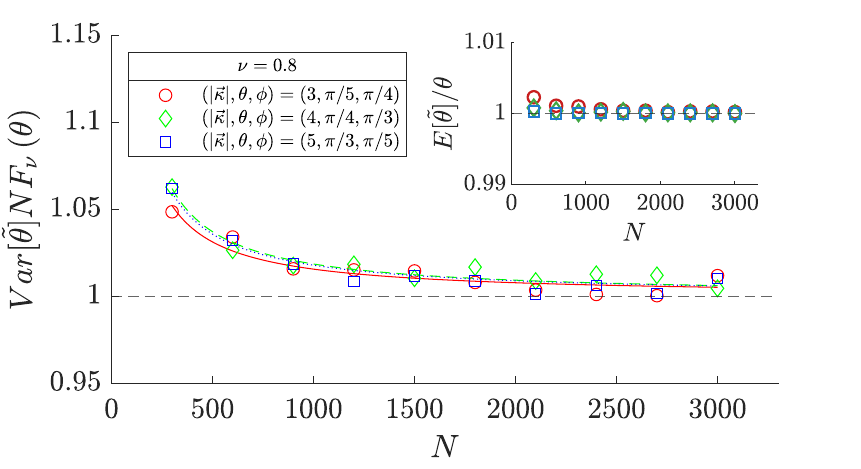}
\includegraphics[width=\linewidth]{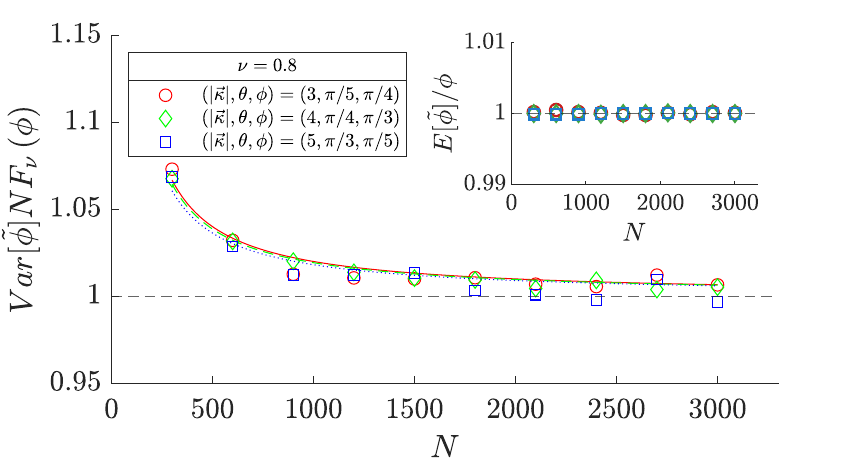}
\end{minipage}%
\caption{Simulations of the variance for the estimation of the parameters $\vert\vec{\kappa}\vert,\theta,\phi$ normalized with respect to the Cram\'{e}r-Rao bound as a function of the number $N$ of experimental iterations for the distinguishability values $\nu=0.7$ (left) and $\nu=0.8$ (right) and the three values of the 3D parameter to estimate: $(\vert\vec{\kappa}\vert,\theta,\phi)=(3,\pi/5,\pi/4),(4,\pi/4,\pi/3),(5,\pi/3,\pi/5)$. In the insets we show the simulations of the expected value of the estimators $(\tilde{\vert\vec{\kappa}\vert},\tilde{\theta},\tilde{\phi})$ normalized to the actual parameter values, showing the unbiasedness of the estimators. We show that the function $1+A/N$ (continuous, dashed and dotted lines, respectively) fit the points of the main figures, where $A$ is the coefficient of the correction term of the order $1/N$ in the variance normalized to the Cram\'{e}r-Rao bound. From the main figure, we can see that the terms of the variance normalized of the order $O(1/N^2)$ are negligible. The term of order $1/N$ corresponds to $1\%$ of the Cram\'{e}r-Rao bound for $N\simeq2000$.}\label{fig:fit}
\end{figure*}

We simulate the sensitivity of this protocol in FIG~\ref{fig:fit}, where we show the variance normalized to the Cram\'{e}r-Rao bound as a function of the size of the sample $N$. The points in the figures are the average of over $100000$ attempted iterations of the likelihood estimators $(\tilde{\vert\vec{\kappa}\vert},\tilde{\theta},\tilde{\phi})$ of the parameters $(\vert\vec{\kappa}\vert,\theta,\phi)$, and they are found numerically by using the set of coupled equations (see~\ref{app:Likelihood})
\begin{equation}
  \displaystyle\sum_i \frac{\alpha\pto{X_i}\nu\sin\pto{\vec{ \rho}_i\cdot\vec{\kappa}}}{1+\alpha\pto{X_i}\nu\cos\pto{\vec{ \rho}_i\cdot\vec{\kappa}}}\vec{ \rho}_i=\begin{pmatrix}
      0\\0\\0
  \end{pmatrix}  ,\label{eq:likelihood3d}
\end{equation}
to solve for $\vec{\kappa}$ as a function of the recorded values $X=X_i$ and $\vec{\rho}=\vec{ \rho}_i$, $i=1,...,N$, at the $i$-th iteration of the experiment. In the inset, we show the expected value of the estimators normalized to the respective value of the parameter to estimate as a function of $N$. Here, we show that the estimators have a bias that is inferior to $1\%$, for any value of $N$, independently of the values of the estimated parameters and of the distinguishability. To analyze the convergence of the precision of our protocol to the Cram\'{e}r-Rao bound, we consider the second-order term of the variance in its asymptotic expansion for large $N$. The variance of the estimators of the parameters $\star=\vert\vec{\kappa}\vert,\theta,\phi$ is approximated as
\begin{eqnarray}
    \eqalign{
        &\mathrm{Var[\tilde{\star}]}\sim\frac{1}{NF_\nu(\star)}+\frac{A}{N^2F_\nu(\star)}\\
        &\Rightarrow\mathrm{Var[\tilde{\star}]}NF_\nu(\star)=1+\frac{A}{N},\label{eq:model}
    }
\end{eqnarray}
which fits the numerical simulations in Fig.~\ref{fig:fit}, for $A\simeq 20$. Therefore, the first order normalized correction of the variance to the Cram\'{e}r-Rao bound is approximately only $0.01$ already for $N\simeq 2000$ iterations.

\section{Optimality even in the case of single-parameter estimation}
In this work, for the very first time, we propose a sensing protocol based on two-photon interference and sampling measurements capable to fully estimate with ultimate quantum precision the difference in all the three components of the momenta of the two photons. In this protocol, measuring all the three localization parameters of the two photons in the near field at the output of the beam splitter increases the indistinguishability of the two photons, enhancing the sensitivity even when the focus is the estimation of only a single parameter.

Let us consider, for example, the probability distribution associated with the case we want to estimate only a single parameter, by measuring only its conjugate parameter at the output of the beam splitter. Taking into account $*=\kappa_x,\kappa_y,\xi$ the general probability distribution is (see~\ref{app:lastlabelipromise})
\begin{equation}
    P_{\nu,\gamma}(X; C^{[*]}|*)=\frac{\gamma^2}{\sqrt{\pi}}\mathrm{e}^{- {C^{[*]}}^2/2}\left(1+\alpha(X)\nu^{[*]}\cos(* C^{[*]})\right),\label{eq: probstarrho}
\end{equation}
where $ C^{[*]}$ is the conjugate variable with respect to $*$, i.e., the variable to which the detectors are sensitive, and the distinguishability is
\begin{equation}
    \nu^{[*]}=\nu\exp\left[\frac{-\vert\vec{\kappa}\vert^2}{2}\right]\exp\left[\frac{*^2}{2}\right].
\end{equation}
Therefore, the Fisher information for the estimation of $*$ is (see~\ref{app:lastlabelipromise})
\begin{equation}
    F_{\nu^{[*]}}(*)=\gamma^2\int \frac{d C^{[*]}}{\sqrt{2\pi}}\mathrm{e}^{- {C^{[*]}}^2/2}\beta_{\nu^{[*]}}(* C^{[*]}) l^2.\label{eq:FIstar1d}
\end{equation}
The likelihood estimator of $*$, i.e., $\tilde{*}$, can be obtained numerically from the equation (see~\ref{app:lastlabelipromise})
\begin{equation}
  \displaystyle\sum_i \frac{\alpha\pto{X_i}\nu^{[*]}\sin\pto{* C^{[*]}_i}}{1+\alpha\pto{X_i}\nu^{[*]}\cos\pto{* C^{[*]}_i}} l_i=0,\label{eq:likelihoodstar}
\end{equation}
to solve for $*$ as a function of the recorded values $X=X_i$ and $C^{[*]}=C^{[*]}_i$, $i=1,...,N$, at the $i$-th iteration of the experiment.

Instead, in the case where we resolve the entire relative three dimensional position vector, the Fisher information matrix for the estimation of  $\kappa_x,\kappa_y,\xi$ is (see~\ref{app:QFINEWstar})
\begin{eqnarray}
    \eqalign{
        F_\nu(\vec{\kappa})&=F_{\nu;I}(\vec{\kappa})I+F_{\nu;\vec{\kappa}}(\vec{\kappa})\frac{\vec{\kappa}\vec{\kappa}^T}{\vert\vec{\kappa}\vert^2}\\&=\frac{\gamma^2}{\sqrt{2\pi}}\int d l \mathrm{e}^{- l^2/2}\beta_\nu(\vert\vec{\kappa}\vert l)\left[I+( l^2-1)\frac{\vec{\kappa}\vec{\kappa}^T}{\vert\vec{\kappa}\vert^2}\right],\label{eq:Fiforcomponents}
    }
\end{eqnarray}
which is the sum of two terms, one proportional to the 3D identity matrix $I$ (we have named its coefficient $F_{\nu;I}(\vec{\kappa})$) and one associated with the projector $\vec{\kappa}\vec{\kappa}^T/\vert\vec{\kappa}\vert^2$ (with coefficient $F_{\nu;\vec{\kappa}}(\vec{\kappa})$). The inverse of this Fisher information matrix is 
\begin{eqnarray}
    \eqalign{
        F_\nu(\vec{\kappa})^{-1}=\frac{1}{F_{\nu;I}(\vec{\kappa})}\left[I-\frac{F_{\nu;\vec{\kappa}}(\vec{\kappa})}{F_{\nu;I}(\vec{\kappa})+F_{\nu;\vec{\kappa}}(\vec{\kappa})}\frac{\vec{\kappa}\vec{\kappa}^T}{\vert\vec{\kappa}\vert^2}\right],\label{eq:fistar3ddritta}
    }
\end{eqnarray}
and its diagonal terms are
\begin{eqnarray}
    \eqalign{
         [F_\nu(\vec{\kappa})^{-1}]_{*,*}=\frac{1}{F_{\nu;I}(\vec{\kappa})}\left[1-\frac{F_{\nu;\vec{\kappa}}(\vec{\kappa})}{F_{\nu;I}(\vec{\kappa})+F_{\nu;\vec{\kappa}}(\vec{\kappa})}\frac{*^2}{\vert\vec{\kappa}\vert^2}\right],
    }\label{eq:FIstar3d}
\end{eqnarray}
which are in general smaller than the inverse of $F_{\nu^{[*]}}(*)$ in Eq.~\eqref{eq:FIstar1d}, i.e. $1/F_{\nu^{[*]}}(*)$, due to the convexity of the Fisher information with respect to marginalization~\cite{zamir2002proof}. Therefore, we have
\begin{equation}
    [F_\nu(\vec{\kappa})^{-1}]_{*,*}\leq  1/F_{\nu^{[*]}}(*).
\end{equation}
Such inequality is saturated only if the only non-zero component of $\vec{\kappa}$ is $*$ ($\nu^{[*]}=\nu$) implying that the Fisher information matrix in Eq.~\eqref{eq:Fiforcomponents} is diagonal with diagonal elements $F_{\nu}(\vec{\kappa})_{*,*}$ equal to $F_\nu(*)$ in Eq.~\eqref{eq:FIstar1d}. However, if any other parameter of $\vec{\kappa}$ than $*$ is non-zero, $\nu^{[*]}<\nu$. Therefore, measuring all three photonic position coordinates in general increases the indistinguishability of the two photons, enhancing the precision in the estimation of any of the three single parameters. In other words, it is in general optimal to estimate the three parameters associated with the photonic momenta through the 3D estimator defined in Eq.~\eqref{eq:likelihood3d} even when interested in only one parameter and not all of them simultaneously. In general, it is always more convenient to achieve indistinguishability in the greatest number of parameters by measuring their conjugate variable, even if some of these parameters are not to be estimated. 

We can also consider the case in which, among all the components of the vector $\vec{\kappa}$, only $*$ is unknown in value, while the other components are known with infinite precision. In this case, the Fisher information matrix can be obtained by considering only one of the diagonal terms of Eq.~\eqref{eq:Fiforcomponents}, obtaining
\begin{equation}
     F_\nu(\vec{\kappa})_{*,*}=F_{\nu;I}(\vec{\kappa})+F_{\nu;\vec{\kappa}}(\vec{\kappa})\frac{*^2}{\vert\vec{\kappa}\vert^2}.\label{eq:fibasta}
\end{equation}
It is possible to show that
\begin{equation}
    1/F_\nu(\vec{\kappa})_{*,*}\leq[F_\nu(\vec{\kappa})^{-1}]_{*,*},
\end{equation}
since knowing with absolute precision the other elements of $\vec{\kappa}$ allows us to estimate with better precision also $*$~\cite{Fiedler1964}. This can be seen by considering the 3D estimator defined by Eq.~\eqref{eq:likelihood3d}, where the three coupled equations do not allow us to estimate separately all the components of $\vec{\kappa}$. Therefore, the uncertainties in all three parameters are strictly connected. Knowing with absolute precision even one parameter- or two, as in the case analyzed in Eq.~\eqref{eq:fibasta}- allows a better estimation of the remaining parameters. 

Therefore, our multi-parameter technique not only provides simultaneous estimation of all three parameters but can also overcome the reduction in sensitivity in previous experiments based on single-parameter estimation (e.g.,\cite{brooks2025quantum,guo2025quantum}). The sensitivity of our multi-parameter estimation technique can be further enhanced using entangled photon pairs achieving a $\sqrt{2}$ improvement per sample~\cite{chen2023spectrally}.

\section{Conclusion}
In this work, we present a three-parameter quantum sensing protocol that can estimate simultaneously and with the ultimate quantum precision the difference in the three components of the momenta of the two photons. We also show that sampling measurements which resolve all the three position parameters of the two photons at the output of the beam splitter increase their indistinguishability, enabling to enhance the sensitivity even when one is interested in the estimation of only a single parameter.

This technique enables us to saturate the quantum Cram\'{e}r-Rao bound for all three photonic momenta parameters simultaneously, with a relatively small number of sampling measurements. Remarkably, the estimators are unbiased for any number $N$ of sampling measurements and the Cram\'{e}r-Rao bound is saturated already for $N\sim 2000$ even for partially distinguishable photons at the detectors. For any number of sampling measurements and values of the parameters to estimate, the bias is always below $1\%$.

Overall, our results place multiparameter quantum-enhanced momentum sensing on firm theoretical ground and indicate a realistic pathway toward scalable, robust quantum metrology protocols operating at the ultimate precision limit in practical free-space quantum technologies, as well as in refractometry and tracking of biological samples.

\section*{Acknowledgments} 
This project is partially supported by Xairos Systems Inc. VT acknowledges support from the Air Force office of Scientific Research under award number FA8655-23-1-7046.
\section*{Data availability statement}
Supporting research data are available on reasonable request from the corresponding author V.T.
\appendix
\section{Evaluation of the output probability distribution in Eq.~\eqref{eq:P1}}\label{app:evolution}
In this section, we evaluate the output probability distribution in Eq.~\eqref{eq:P1}.
The channels 1 and 2 are the input port of a balanced beam splitter, whose effect on the state $\ket{\Phi}$ in Eq.~\eqref{maineq:input} is described by the operator $\hat{U}_{BS}$. This operator acts on the injected probe through the map $\hat{U}_{BS}\hat{a}^\dagger_i\hat{U}_{BS}=\sum_j U_{BS,ij}\hat{a}^\dagger_j$, and $\hat{U}_{BS}\hat{b}^\dagger_i\hat{U}_{BS}=\sum_j U_{BS,ij}\hat{b}^\dagger_j$, where
\begin{equation}
    U_{BS}=\frac{1}{\sqrt{2}}\begin{pmatrix}
        1&-1\\1&1
    \end{pmatrix}.\label{appeq:BS}
\end{equation}
By using Eq.~\eqref{appeq:BS} and Eq.~\eqref{maineq:input},  the two-photon state after the beam splitter is
\begin{eqnarray}
\eqalign{
    \ket{\Phi^{\mathrm{BS}}}=\hat{U}_{BS}\ket{\Phi}&=\int d^3\vec{r}_1 \psi_1\pto{\vec{r}_1}\frac{\hat{a}^\dagger_1\pto{\vec{r}_1}-\hat{a}^\dagger_2\pto{\vec{r}_1}}{\sqrt{2}}\ket{0}\\&\otimes\int d^3\vec{r}_2 \psi_2\pto{\vec{r}_2}\frac{\hat{d}^\dagger_1\pto{\vec{r}_2}+\hat{d}^\dagger_2\pto{\vec{r}_2}}{\sqrt{2}}\ket{0}.}
\end{eqnarray}
This state is defined by three contributes. In fact, we can write
\begin{eqnarray}
    \eqalign{
        \ket{\Phi^{\mathrm{BS}}}&=\frac{1}{2}\int d^3\vec{r}_1 d\vec{r}_2\psi_1\pto{\vec{r}_1}\psi_2\pto{\vec{r}_2}\hat{a}^\dagger_1\pto{\vec{r}_1}\hat{d}^\dagger_1\pto{\vec{r}_2}\ket{0}+\\
        &-\frac{1}{2}\int d^3\vec{r}_1 d\vec{r}_2\psi_1\pto{\vec{r}_1}\psi_2\pto{\vec{r}_2}\hat{a}^\dagger_2\pto{\vec{r}_1}\hat{d}^\dagger_2\pto{\vec{r}_2}\ket{0}+\\
        &+\frac{1}{2}\int d^3\vec{r}_1 d\vec{r}_2\psi_1\pto{\vec{r}_1}\psi_2\pto{\vec{r}_2}\pto{\hat{a}^\dagger_1\pto{\vec{r}_1}\hat{d}^\dagger_2\pto{\vec{r}_2}-\hat{a}^\dagger_2\pto{\vec{r}_1}\hat{d}^\dagger_1\pto{\vec{r}_2}}\ket{0},\label{eq:finalstate}
    }
\end{eqnarray}
where the first and second rows represent, respectively, a bunching (B) on output channels 1 and 2, and the third row represents an anti-bunching (A).

We can evaluate the probability of having an anti-bunching event while measuring $\pto{\vec{r}_1,\vec{r}_2}$ using Eq.~\eqref{eq:finalstate},
\begin{eqnarray}
    \eqalign{
        P\pto{A;\vec{r}_1,\vec{r}_2\vert \vec{k}_1,\vec{k}_2}&=\left\vert \bra{0}\hat{a}_1\pto{\vec{r}_1}\hat{a}_2\pto{\vec{r}_2}\ket{\Phi^{\mathrm{BS}}}\right\vert^2+\\
        &+\left\vert \bra{0}\hat{a}_1\pto{\vec{r}_1}\hat{b}_2\pto{\vec{r}_2}\ket{\Phi^{\mathrm{BS}}}\right\vert^2+\\&+\left\vert \bra{0}\hat{b}_1\pto{\vec{r}_1}\hat{a}_2\pto{\vec{r}_2}\ket{\Phi^{\mathrm{BS}}}\right\vert^2+\\&+\left\vert \bra{0}\hat{b}_1\pto{\vec{r}_1}\hat{b}_2\pto{\vec{r}_2}\ket{\Phi^{\mathrm{BS}}}\right\vert^2.\label{eq:PAdef}
    }
\end{eqnarray}
The first term reads
\begin{eqnarray}
    \eqalign{
      \left\vert \bra{0}\hat{a}_1\pto{\vec{r}_1}\hat{a}_2\pto{\vec{r}_2}\ket{\Phi^{\mathrm{BS}}}\right\vert^2&=\left\vert\frac{\sqrt{\nu}}{2}\pto{\psi_2\pto{\vec{r}_2}\psi_1\pto{\vec{r}_1}-\psi_2\pto{\vec{r}_1}\psi_1\pto{\vec{r}_2}} \right\vert^2 \\
     &=\frac{\nu}{4}[\left\vert\psi_2\pto{\vec{r}_2}\psi_1\pto{\vec{r}_1}\right\vert^2+\left\vert\psi_2\pto{\vec{r}_1}\psi_1\pto{\vec{r}_2}\right\vert^2\\&-2\mathrm{Re}\left(\psi_2\pto{\vec{r}_2}\psi_1\pto{\vec{r}_1}\psi_2^{*}\pto{\vec{r}_1}\psi_1^{*}\pto{\vec{r}_2}\right)].\label{eq:PA1}
    }
\end{eqnarray}
We notice that this term already has a term of interference in the 3D spatial domain. This is the only term defining Eq.~\eqref{eq:PAdef} that has this property, since it is the only term in which the optical mode is the same for both detected photons (and identified by $\hat{a}$). In the other non-null terms, the two optical modes are defined by $\hat{a}$ and $\hat{b}$. These other terms are
\begin{eqnarray}
    \eqalign{
        \left\vert \bra{0}\hat{a}_1\pto{\vec{r}_1}\hat{b}_2\pto{\vec{r}_2}\ket{\Phi^{\mathrm{BS}}}\right\vert^2&=\left\vert\frac{\sqrt{1-\nu}}{2}\psi_2\pto{\vec{r}_2}\psi_1\pto{\vec{r}_1}\right\vert^2\\
        &=\frac{1-\nu}{4}\left\vert \psi_1\pto{\vec{r}_1} \psi_2\pto{\vec{r}_2}\right\vert^2,\label{eq:PA2}
    }
\end{eqnarray}
\begin{eqnarray}
    \eqalign{
        \left\vert \bra{0}\hat{b}_1\pto{\vec{r}_1}\hat{a}_2\pto{\vec{r}_2}\ket{\Phi^{\mathrm{BS}}}\right\vert^2&=\left\vert\frac{\sqrt{1-\nu}}{2}\psi_1\pto{\vec{r}_2}\psi_2\pto{\vec{r}_1}\right\vert^2\\
        &=\frac{1-\nu}{4}\left\vert \psi_2\pto{\vec{r}_1} \psi_1\pto{\vec{r}_2}\right\vert^2.\label{eq:PA3}
    }
\end{eqnarray}
The only null term is the one in which both the detected photons are in the optical mode $\hat{b}$, which is impossible since the optical mode of the photon entering the input channel 1 is $\hat{a}$. Therefore, 
\begin{equation}
    \left\vert \bra{0}\hat{b}_1\pto{\vec{r}_1}\hat{b}_2\pto{\vec{r}_2}\ket{\Phi^{\mathrm{BS}}}\right\vert^2=0.\label{eq:PA4}
\end{equation}
The sum of Eqs.~\eqref{eq:PA1},~\eqref{eq:PA2},\eqref{eq:PA3},\eqref{eq:PA4} allows to write Eq.~\eqref{eq:PAdef} as follows
\begin{eqnarray}
    \eqalign{
     P\pto{A;\vec{r}_1,\vec{r}_2\vert \vec{k}_1,\vec{k}_2}&=\frac{1}{4}[\left\vert \psi_1\pto{\vec{r}_1} \psi_2\pto{\vec{r}_2} \right\vert^2+\left\vert \psi_2\pto{\vec{r}_1} \psi_1\pto{\vec{r}_2} \right\vert^2\\&-2\nu\mathrm{Re}\left(\psi_2\pto{\vec{r}_2}\psi_1\pto{\vec{r}_1}\psi_2^{*}\pto{\vec{r}_1}\psi_1^{*}\pto{\vec{r}_2}\right)].}\label{Eq:PA}
\end{eqnarray}

Instead, the probability of having a bunching event, while measuring $\pto{\vec{r}_1,\vec{r}_2}$, is
\begin{eqnarray}
    \eqalign{
        P\pto{B;\vec{r}_1,\vec{r}_2\vert \vec{k}_1,\vec{k}_2}&=\frac{1}{2}\sum_{i=1,2}\left\vert \bra{0}\hat{a}_i\pto{\vec{r}_1}\hat{a}_i\pto{\vec{r}_2}\ket{\Phi^{\mathrm{BS}}}\right\vert^2+\\
        &+\sum_{i=1,2}\left\vert \bra{0}\hat{a}_i\pto{\vec{r}_1}\hat{b}_i\pto{\vec{r}_2}\ket{\Phi^{\mathrm{BS}}}\right\vert^2+\\&+\frac{1}{2}\sum_{i=1,2}\left\vert \bra{0}\hat{b}_i\pto{\vec{r}_1}\hat{b}_i\pto{\vec{r}_2}\ket{\Phi^{\mathrm{BS}}}\right\vert^2.
    }
\end{eqnarray}
The first term is
\begin{eqnarray}
    \eqalign{
        &\frac{1}{2}\sum_{i=1,2}\left\vert \bra{0}\hat{a}_i\pto{\vec{r}_1}\hat{a}_i\pto{\vec{r}_2}\ket{\Phi^{\mathrm{BS}}}\right\vert^2\\&=\left\vert\frac{\sqrt{\nu}}{2}\pto{\psi_2\pto{\vec{r}_2}\psi_1\pto{\vec{r}_1}+\psi_2\pto{\vec{r}_1}\psi_1\pto{\vec{r}_2}} \right\vert^2 \\
       &=\frac{\nu}{4}[\left\vert\psi_2\pto{\vec{r}_2}\psi_1\pto{\vec{r}_1}\right\vert^2+\left\vert\psi_2\pto{\vec{r}_1}\psi_1\pto{\vec{r}_2}\right\vert^2\\&+2\mathrm{Re}\left(\psi_2\pto{\vec{r}_2}\psi_1\pto{\vec{r}_1}\psi_2^{*}\pto{\vec{r}_1}\psi_1^{*}\pto{\vec{r}_2}\right)],
    }
\end{eqnarray}
which present an interference effect in the 3D spatial domain like the one in Eq.~\eqref{eq:PA1}. Instead, the second term does not have any interference in the 3D spatial domain, like in Eqs.~\eqref{eq:PA2},\eqref{eq:PA3}, as we can see below 
\begin{eqnarray}
    \sum_{i=1,2}\left\vert \bra{0}\hat{a}_i\pto{\vec{r}_1}\hat{b}_i\pto{\vec{r}_2}\ket{\Phi^{\mathrm{BS}}}\right\vert^2
        &=\frac{1-\nu}{4}[\left\vert\psi_2\pto{\vec{r}_2}\psi_1\pto{\vec{r}_1}\right\vert^2\\&+\left\vert\psi_2\pto{\vec{r}_1}\psi_1\pto{\vec{r}_2}\right\vert^2].
\end{eqnarray}
The third is null, for the same reason for which Eq.~\eqref{eq:PA4} is null, 
\begin{eqnarray}
    \eqalign{
        \frac{1}{2}\sum_{i=1,2}\left\vert \bra{0}\hat{b}_i\pto{\vec{r}_1}\hat{b}_i\pto{\vec{r}_2}\ket{\Phi^{\mathrm{BS}}}\right\vert^2=0.
    }
\end{eqnarray}
Therefore, the bunching probability is
\begin{eqnarray}
\eqalign{
     P\pto{B;\vec{r}_1,\vec{r}_2\vert \vec{k}_1,\vec{k}_2}&=\frac{1}{4}[\left\vert \psi_1\pto{\vec{r}_1} \psi_2\pto{\vec{r}_2} \right\vert^2+\left\vert \psi_2\pto{\vec{r}_1} \psi_1\pto{\vec{r}_2} \right\vert^2\\&+2\nu\mathrm{Re}\left(\psi_2\pto{\vec{r}_2}\psi_1\pto{\vec{r}_1}\psi_2^{*}\pto{\vec{r}_1}\psi_1^{*}\pto{\vec{r}_2}\right)].
     }\label{Eq:PB}
\end{eqnarray}

We notice that Eq.~\eqref{Eq:PA} and Eq.~\eqref{Eq:PB} are different only by a change of sign of the last term. Therefore, we define $\alpha(X)$, where $\alpha(A)=-1$, and $\alpha(B)=1$, with whom it is possible to write Eq.~\eqref{Eq:PA} and Eq.~\eqref{Eq:PB} as follows
\begin{eqnarray}
\eqalign{
    P\pto{X;\vec{r}_1,\vec{r}_2\vert \vec{k}_1,\vec{k}_2}&=\frac{1}{4}[\left\vert \psi_1\pto{\vec{r}_1} \psi_2\pto{\vec{r}_2} \right\vert^2+\left\vert \psi_2\pto{\vec{r}_1} \psi_1\pto{\vec{r}_2} \right\vert^2\\&+2\alpha(X)\nu\mathrm{Re}\left(\psi_2\pto{\vec{r}_2}\psi_1\pto{\vec{r}_1}\psi_2^{*}\pto{\vec{r}_1}\psi_1^{*}\pto{\vec{r}_2}\right)].}\label{app:prop(psi)}
\end{eqnarray}
Using Eq.~\eqref{maineq:amplt}, it is possible to notice that the beating term is only function of the difference between $\vec{r}_2$ and $\vec{r}_1$ and the difference between $\vec{k}_2$ and $\vec{k}_1$, namely $\Delta \vec{r}$ and $\Delta \vec{k}$
\begin{equation}
  \mathrm{Re}\left(\psi_2\pto{\vec{r}_2}\psi_1\pto{\vec{r}_1}\psi_2^{*}\pto{\vec{r}_1}\psi_1^{*}\pto{\vec{r}_2}\right)= \left\vert\psi_2\pto{\vec{r}_2}\psi_1\pto{\vec{r}_1}\psi_2\pto{\vec{r}_1}\psi_1\pto{\vec{r}_2}\right\vert \cos\left(\Delta \vec{r}\cdot\Delta \vec{k}\right).
\end{equation}
Therefore, in order to detect the beating, we can just be sensitive to the difference between the localization parameters. Defining 
\begin{equation}
    \vec{r}_m=\frac{\vec{r}_1+\vec{r}_2}{2}\qquad ,\qquad \Delta\vec{r}=\vec{r}_2-\vec{r}_1,\label{app:m}
\end{equation}
and
\begin{equation}
    \vec{k}_m=\frac{\vec{k}_1+\vec{k}_2}{2}\qquad ,\qquad \Delta\vec{k}=\vec{k}_2-\vec{k}_1,\label{app:delta}
\end{equation}
we can parameterize
$\vec{r}_2=\vec{r}_m+\Delta \vec{r}/2$ and $\vec{k}_2=\vec{k}_m+\Delta \vec{k}/2$. Integrating Eq.~\eqref{app:prop(psi)}, we can obtain the output probabilities for the case in which we measure at the output of the beam splitter only $\Delta\vec{r}$
\begin{eqnarray}
    \eqalign{
         P_\nu\left(X;\Delta \vec{r}|\Delta \vec{k}\right)&=\int dr_1P\pto{X;\vec{r}_1,\vec{r}_2\vert \vec{k}_1,\vec{k}_2}=\frac{1}{\sqrt{(2\pi)^{3}\det(\Sigma_1+\Sigma_2)}}\times\\
        &\times\frac{1}{2}\Bigg\lbrace\exp\left[-\Delta\vec{r}^T\frac{(\Sigma_1+\Sigma_2)^{-1}}{2}\Delta\vec{r}\right]\\
        &+\alpha(X)\nu\exp\left[-\Delta\vec{r}^T\frac{\Sigma^{-1}_1+\Sigma^{-1}_2}{8}\Delta\vec{r}\right]\cos\left(\Delta \vec{r}\cdot\Delta \vec{k}\right)\Bigg\rbrace.
    }
\end{eqnarray}

\begin{eqnarray}
    \eqalign{
         P_\nu\left(X;\Delta \vec{r}|\Delta \vec{k}\right)&=\frac{1}{4\sqrt{(2\pi)^{3}\det(\Sigma_1+\Sigma_2)}}\exp\left[-\Delta\vec{r}^T\frac{(\Sigma_1+\Sigma_2)^{-1}}{2}\Delta\vec{r}\right]\\
        &\times\left[1+\alpha(X)\nu\mathrm{e}^{-\Delta\vec{r}^T\left(\frac{\Sigma^{-1}_1+\Sigma^{-1}_2}{8}-\frac{(\Sigma_1+\Sigma_2)^{-1}}{2}\right)\Delta\vec{r}}\cos\left(\Delta \vec{r}\cdot\Delta \vec{k}\right)\right].
    }
\end{eqnarray}
The term $\nu\exp\left[-\Delta\vec{r}^T\left(\frac{\Sigma^{-1}_1+\Sigma^{-1}_2}{8}-\frac{(\Sigma_1+\Sigma_2)^{-1}}{2}\right)\Delta\vec{r}\right]$ represents the visibility of the beatings. In order to minimize it, me must impose $\Sigma_1=\Sigma_2$. However, this might be experimentally challenging to reproduce. However, we can prove that if the covariance matrices can be slightly different between each others in order to maximize the variance. In order to do that, we impose $\delta\Sigma = \Sigma_2 - \Sigma_1$, with $\delta\Sigma \ll \Sigma_1, \Sigma_2$. Here, inequalities are understood in terms of the matrix norm~\cite{kato2013perturbation}. Therefore, by imposing the parametrization $\Sigma_2=\Sigma_1+\delta \Sigma$, we have $\Sigma_2^{-1}=\Sigma_1^{-1}-\Sigma_1^{-1}\delta \Sigma\Sigma_1^{-1} +O(\delta\Sigma^2)$ and $(\Sigma_1+\Sigma_2)^{-1}=\Sigma_1^{-1}/2-\Sigma_1^{-1}\delta \Sigma\Sigma_1^{-1}/4 +O(\delta\Sigma^2)$. This leads to
\begin{equation}
    \frac{\Sigma^{-1}_1+\Sigma^{-1}_2}{8}-\frac{(\Sigma_1+\Sigma_2)^{-1}}{2}=O(\delta\Sigma^2),
\end{equation}
which is a negligible term. If we consider this, the output probability can be further simplified 
\begin{eqnarray}
    \eqalign{
          P_\nu\left(X;\Delta \vec{r}|\Delta \vec{k}\right)&=\frac{1}{\sqrt{(2\pi)^{3}\det(\Sigma)}}\exp\left[-\Delta\vec{r}^T\frac{\Sigma^{-1}}{2}\Delta\vec{r}\right]\times\\
        &\times\frac{1}{2}\Bigg\lbrace1+\alpha(X)\nu\cos\left(\Delta \vec{r}\cdot\Delta \vec{k}\right)\Bigg\rbrace,
    }
\end{eqnarray}
where $\Sigma=\Sigma_1+\Sigma_2=2\Sigma_1+\delta\Sigma$, with eigenvalues $\sigma^2_k$, where $k=x,y,ct$. We consider now the presence of losses in the detection. We assume that both the efficiencies of the two detectors are equal to $\gamma\in\pq{0,1}$. Each photon can be lost in the detection with probability $1-\gamma$. Therefore, the probabilities of detecting zero, one, or two photons are, respectively, 
\begin{eqnarray}
    \eqalign{
       & P_0=\pto{1-\gamma}^2,\\
       & P_1=\gamma\pto{1-\gamma}\int dr_1 \sum_X P_\nu(X;r_1,r_2|k_1,k_2)=2\gamma(1-\gamma),\\
       & P_{\nu,\gamma}\pto{X;\Delta\vec{r}\vert  \Delta\vec{k}}=\gamma^2P_\nu\left(X;\Delta\vec{r}\vert  \Delta\vec{k}\right).\label{app:resprob}
    }
\end{eqnarray}
The zero-photon and one-photon events do not carry any information for the estimation of $\Delta \vec{k}$. In fact, only the two-photon events have a probability that is function of these two parameters. Therefore, from now on, we will consider, instead of the whole set of probabilities, only $ P_{\nu,\gamma}\pto{X;\Delta\vec{r}\vert  \Delta\vec{k}}$. The result is
\begin{eqnarray}
    \eqalign{
          P_{\nu,\gamma}\left(X;\Delta \vec{r}|\Delta \vec{k}\right)&=\frac{\gamma^2}{\sqrt{(2\pi)^{3}\det(\Sigma)}}\exp\left[-\Delta\vec{r}^T\frac{\Sigma^{-1}}{2}\Delta\vec{r}\right]\times\\
        &\times\frac{1}{2}\Bigg\lbrace1+\alpha(X)\nu\cos\left(\Delta \vec{r}\cdot\Delta \vec{k}\right)\Bigg\rbrace,
    }
\end{eqnarray}
which is Eq.~\eqref{eq:P1} in the main text.
\section{Verification of the conditions of saturability of the quantum Cram\'{e}r-Rao bound in Eq.~\eqref{eq:SLD}}\label{SLD}
In this section, we verify that Eq.~\eqref{eq:SLD} is obtained for the symmetric logarithmic derivatives for the parameters $\vert\vec{\kappa}\vert,\theta,\phi$ in Eq.~\eqref{app:defSLD}. From Eq.~\eqref{app:defSLD}, we can find the expected values of the commutators of the symmetric logarithmic derivatives, which are
\begin{eqnarray}
    \eqalign{
        &\mathrm{Tr\left[\ket{\Phi}\bra{\Phi}\left[\hat{L}_{\vert\vec{\kappa}\vert},\hat{L}_\theta\right]\right]}=8\,\mathrm{Im}\left[\Braket{\frac{\partial\Phi}{\partial \vert\vec{\kappa}\vert}}{\Phi}\Braket{\Phi}{\frac{\partial\Phi}{\partial\theta}}-\Braket{\frac{\partial\Phi}{\partial \vert\vec{\kappa}\vert}}{\frac{\partial\Phi}{\partial\theta}}\right],\\
        &\mathrm{Tr\left[\ket{\Phi}\bra{\Phi}\left[\hat{L}_{\theta},\hat{L}_\phi\right]\right]}=8\,\mathrm{Im}\left[\Braket{\frac{\partial\Phi}{\partial \theta}}{\Phi}\Braket{\Phi}{\frac{\partial\Phi}{\partial\phi}}-\Braket{\frac{\partial\Phi}{\partial \theta}}{\frac{\partial\Phi}{\partial\phi}}\right],\\
        &\mathrm{Tr\left[\ket{\Phi}\bra{\Phi}\left[\hat{L}_{\phi},\hat{L}_\vert\vec{\kappa}\vert\right]\right]}=8\,\mathrm{Im}\left[\Braket{\frac{\partial\Phi}{\partial \phi}}{\Phi}\Braket{\Phi}{\frac{\partial\Phi}{\partial \vert\vec{\kappa}\vert}}-\Braket{\frac{\partial\Phi}{\partial \phi}}{\frac{\partial\Phi}{\partial \vert\vec{\kappa}\vert}}\right].\\
    }\label{app:8Im}
\end{eqnarray}
In order to derive all the terms in Eq.~\eqref{app:8Im}, we remind from Eq.~\eqref{maineq:input} that
\begin{eqnarray}
    \eqalign{
        \ket{\Phi}&=\int d^3r_1d^3r_2 \mathrm{e}^{-i\vec{r}_1\cdot\vec{k}_1-i\vec{r}_2\cdot\vec{k}_2}\left\vert\psi_1\pto{\vec{r}_1}\psi_2\pto{\vec{r}_2}\right\vert\hat{a}^\dagger_1\pto{\vec{r}_1}\hat{d}^\dagger_2\pto{\vec{r}_2}\ket{0}.\label{app:sldPhi}
    }
\end{eqnarray}
Here, only the phase depends on $\vec{k}_j$, $j=1,2$, therefore, only the phase will depend on $\vert\vec{\kappa}\vert,\theta,\phi$. We can manipulate the argument of the phase in order to explicit the dependence on these parameters
\begin{eqnarray}
    \eqalign{\vec{r}_1\cdot\vec{k}_1+\vec{r}_2\cdot\vec{k}_2&=\frac{1}{2}\Delta\vec{r}\cdot\Delta\vec{k}+\vec{r}_{m}\cdot\vec{k}_m\\
    &=\frac{1}{2}\vec{\kappa}\cdot\vec{ \rho}+\vec{r}_m\cdot\vec{k}_m,
    }
\end{eqnarray}
with $\Delta\vec{r}$,$\Delta \vec{k}$, $\vec{r}_m$, $\vec{k}_m$ defined in Eq.~\eqref{app:delta} and Eq.~\eqref{app:m}, while $\vec{\kappa}$ and $\vec{\rho}$ are defined in Eq.~\eqref{eq:reparameters}.
Therefore, we have
\begin{eqnarray}
    \eqalign{
    \ket{\frac{\partial\Phi}{\partial \vert\vec{\kappa}\vert}}&=-\frac{i}{2}\int d^3r_1d^3r_2 \frac{\vec{\kappa}\cdot\vec{ \rho}}{\vert\vec{\kappa}\vert}\mathrm{e}^{-i\vec{r}_1\cdot\vec{k}_1-i\vec{r}_2\cdot\vec{k}_2}\\&\times\left\vert\psi_1\pto{\vec{r}_1}\psi_2\pto{\vec{r}_2}\right\vert\hat{a}^\dagger_1\pto{\vec{r}_1}\hat{d}^\dagger_2\pto{\vec{r}_2}\ket{0},\\
        \ket{\frac{\partial\Phi}{\partial \theta}}&=-\frac{i}{2}\int d^3r_1d^3r_2 \frac{\partial\vec{\kappa}}{\partial\theta}\cdot\vec{ \rho}\mathrm{e}^{-i\vec{r}_1\cdot\vec{k}_1-i\vec{r}_2\cdot\vec{k}_2}\\&\times\left\vert\psi_1\pto{\vec{r}_1}\psi_2\pto{\vec{r}_2}\right\vert\hat{a}^\dagger_1\pto{\vec{r}_1}\hat{d}^\dagger_2\pto{\vec{r}_2}\ket{0},\\
        \ket{\frac{\partial\Phi}{\partial \phi}}&=-\frac{i}{2}\int d^3r_1d^3r_2 \frac{\partial\vec{\kappa}}{\partial\phi}\cdot\vec{ \rho}\mathrm{e}^{-i\vec{r}_1\cdot\vec{k}_1-i\vec{r}_2\cdot\vec{k}_2}\\&\times\left\vert\psi_1\pto{\vec{r}_1}\psi_2\pto{\vec{r}_2}\right\vert\hat{a}^\dagger_1\pto{\vec{r}_1}\hat{d}^\dagger_2\pto{\vec{r}_2}\ket{0}.\label{app:sldpartialPhi}
    }
\end{eqnarray}
From here, it is easy to show that
\begin{eqnarray}
\eqalign{
    \Braket{\frac{\partial\Phi}{\partial \vert\vec{\kappa}\vert}}{\frac{\partial\Phi}{\partial \vert\vec{\kappa}\vert}}&=\frac{1}{4(2\pi)^{3/2}}\int d^3 \rho\mathrm{e}^{-\vert\vec{ \rho}\vert^2/2}
       \frac{(\vec{\kappa}\cdot\vec{ \rho})^2}{\vert\vec{\kappa}\vert^2}=\frac{1}{4},\\
    \Braket{\frac{\partial\Phi}{\partial \theta}}{\frac{\partial\Phi}{\partial \theta}}&=\frac{1}{4(2\pi)^{3/2}}\int d^3 \rho\mathrm{e}^{-\vert\vec{ \rho}\vert^2/2}
       \left(\vec{ \rho}\cdot\frac{\partial\vec\kappa}{\partial\theta}\right)^2=\frac{\vert\vec{\kappa}\vert^2}{4},\\
    \Braket{\frac{\partial\Phi}{\partial \phi}}{\frac{\partial\Phi}{\partial \phi}}&=\frac{1}{4(2\pi)^{3/2}}\int d^3\rho\mathrm{e}^{-\vert\vec{ \rho}\vert^2/2}
       \left(\vec{ \rho}\cdot\frac{\partial\vec\kappa}{\partial\phi}\right)^2=\frac{\vert\vec{\kappa}\vert^2\sin^2\theta}{4},\label{app:derbraderket}
       }
\end{eqnarray}
and
\begin{eqnarray}
\eqalign{
\Braket{\frac{\partial\Phi}{\partial \vert\vec{\kappa}\vert}}{\frac{\partial\Phi}{\partial \theta}}&=\frac{1}{4(2\pi)^{3/2}}\int d^3 \rho\mathrm{e}^{-\vert\vec{ \rho}\vert^2/2}
       \frac{(\vec{\kappa}\cdot\vec{ \rho})}{\vert\vec{\kappa}\vert}\left(\vec{ \rho}\cdot\frac{\partial\vec\kappa}{\partial\theta}\right)=0,\\
    \Braket{\frac{\partial\Phi}{\partial \theta}}{\frac{\partial\Phi}{\partial \phi}}&=\frac{1}{4(2\pi)^{3/2}}\int d^3 \rho\mathrm{e}^{-\vert\vec{ \rho}\vert^2/2}
       \left(\vec{ \rho}\cdot\frac{\partial\vec\kappa}{\partial\theta}\right)\left(\vec{ \rho}\cdot\frac{\partial\vec\kappa}{\partial\phi}\right)=0,\\
    \Braket{\frac{\partial\Phi}{\partial \phi}}{\frac{\partial\Phi}{\partial \vert\vec{\kappa}\vert}}&=\frac{1}{4(2\pi)^{3/2}}\int d^3\rho\mathrm{e}^{-\vert\vec{ \rho}\vert^2/2}
       \left(\vec{ \rho}\cdot\frac{\partial\vec\kappa}{\partial\phi}\right)\frac{(\vec{\kappa}\cdot\vec{ \rho})}{\vert\vec{\kappa}\vert}=0,\label{app:derbraderketoffdiagonal}}
\end{eqnarray}
Also,
\begin{eqnarray}
    \eqalign{
      \Braket{\Phi}{\frac{\partial\Phi}{\partial s}}&=-\frac{i}{2}\int d^3r_1d^3r_2 \frac{\vec{\kappa}\cdot\vec{ \rho}}{\vert\vec{\kappa}\vert}\left\vert\psi_1\pto{\vec{r}_1}\psi_2\pto{\vec{r}_2}\right\vert^2,\\\Braket{\Phi}{\frac{\partial\Phi}{\partial \theta}}&=-\frac{i}{2}\int  d^3r_1d^3r_2 \vec{ \rho}\cdot\frac{\partial\vec{\kappa}}{\partial\theta}\left\vert\psi_1\pto{\vec{r}_1}\psi_2\pto{\vec{r}_2}\right\vert^2,\\\Braket{\Phi}{\frac{\partial\Phi}{\partial \phi}}&=-\frac{i}{2}\int  d^3r_1d^3r_2 \vec{ \rho}\cdot\frac{\partial\vec{\kappa}}{\partial\phi}\left\vert\psi_1\pto{\vec{r}_1}\psi_2\pto{\vec{r}_2}\right\vert^2.\label{app:brap1ketp2}
    }
\end{eqnarray}
which leads to
\begin{eqnarray}
    \eqalign{
      \Braket{\Phi}{\frac{\partial\Phi}{\partial \vert\vec{\kappa}\vert}}&=-\frac{i}{2(2\pi)^{3/2}}\int d^3 \rho\mathrm{e^{-\vert \rho\vert^2/2}} \frac{\vec{\kappa}\cdot\vec{ \rho}}{\vert\vec{\kappa}\vert}=0,\\\Braket{\Phi}{\frac{\partial\Phi}{\partial \theta}}&=-\frac{i}{2(2\pi)^{3/2}}\int d^3 \rho\mathrm{e^{-\vert \rho\vert^2/2}} \vec{ \rho}\cdot\frac{\partial\vec{\kappa}}{\partial\theta}=0,\\\Braket{\Phi}{\frac{\partial\Phi}{\partial \phi}}&=-\frac{i}{2(2\pi)^{3/2}}\int d^3 \rho\mathrm{e^{-\vert \rho\vert^2/2}} \vec{ \rho}\cdot\frac{\partial\vec{\kappa}}{\partial\phi}=0.\label{app:referee}
    }
\end{eqnarray}

Using Eq.~\eqref{app:brap1ketp2} and Eq.~\eqref{app:referee}, we can prove that
\begin{eqnarray}
    \eqalign{
        &\mathrm{Tr\left[\ket{\Phi}\bra{\Phi}\left[\hat{L}_{\vert\vec{\kappa}\vert},\hat{L}_\theta\right]\right]}=0,\\
        &\mathrm{Tr\left[\ket{\Phi}\bra{\Phi}\left[\hat{L}_{\theta},\hat{L}_\phi\right]\right]}=0,\\
        &\mathrm{Tr\left[\ket{\Phi}\bra{\Phi}\left[\hat{L}_{\phi},\hat{L}_{\vert\vec{\kappa}\vert}\right]\right]}=0,\\ \label{app:SLD}
    }
\end{eqnarray}
which is Eq.~\eqref{eq:SLD} of the main text.

\section{Evaluation of the quantum Fisher information in Eq.\eqref{eq:Q}}
\label{app:QFI}
In this section, we evaluate the quantum Fisher information matrix in Eq.\eqref{eq:Q}, which will assume the form
\begin{equation}
    Q(\vert\vec{\kappa}\vert,\theta,\phi)=\begin{pmatrix}
        Q_{\vert\vec{\kappa}\vert\vert\vec{\kappa}\vert}&Q_{\vert\vec{\kappa}\vert\theta}&Q_{\vert\vec{\kappa}\vert\phi}\\
        Q_{\vert\vec{\kappa}\vert\theta}&Q_{\theta\theta}&Q_{\theta\phi}\\
        Q_{\vert\vec{\kappa}\vert\phi}&Q_{\theta\phi}&Q_{\phi\phi}
    \end{pmatrix}\label{appeq:Q}
    ,
\end{equation}
where, the elements of the matrix are by definition
\begin{eqnarray}
    \eqalign{
Q_{\zeta\mu}=4\mathrm{Re}\pq{\Braket{\frac{\partial\Phi}{\partial\zeta}}{\frac{\partial\Phi}{\partial\mu}}-\Braket{\frac{\partial\Phi}{\partial\zeta}}{\Phi}\Braket{\Phi}{\frac{\partial\Phi}{\partial\mu}}},
    }\label{app:Qzm}
\end{eqnarray}
where $\zeta,\mu=\vert\vec{\kappa}\vert,\theta,\phi$. We remind that $\ket{\Phi}$ is defined in Eq.~\eqref{app:sldPhi} and $\ket{\partial \Phi/\partial \mu}$, $\mu=\vert\vec{\kappa}\vert,\theta,\phi$ is defined in Eq.~\eqref{app:sldpartialPhi}. The inner products that define the quantum Fisher information matrix have been evaluated in Eq.~\eqref{app:derbraderket}, Eq.~\eqref{app:derbraderketoffdiagonal} and Eq.~\eqref{app:referee}. Using these equations, we obtain
\begin{equation}
    Q(\vert\vec{\kappa}\vert,\theta,\phi)=\begin{pmatrix}
        1&0&0\\0&\vert\vec{\kappa}\vert^2&0\\0&0&\vert\vec{\kappa}\vert^2\sin^2\theta
    \end{pmatrix},
\end{equation}
which is the Eq.~\eqref{eq:Q} in the main text.

\section{Evaluation of the Fisher information matrix in Eqs.~\eqref{eq:finu1},\eqref{eq:Fi} under the condition of maximum and partial visibility, respectively }
\label{app:FI}
In this section, we will analyze in detail the Fisher information matrix for the sensing scheme, its density, and the Fisher information matrix for the non-resolving protocol. 

By using the probability distribution in Eq.~\eqref{maineq:prob}, the Fisher information matrix for the parameters $\vert\vec{\kappa}\vert,\theta,\phi$ is the following
\begin{eqnarray}
\eqalign{
    F_\nu(\vert\vec{\kappa}\vert,\theta,\phi)&=\begin{pmatrix}
        F_{\vert\vec{\kappa}\vert\vert\vec{\kappa}\vert}&F_{\vert\vec{\kappa}\vert\theta}&F_{\vert\vec{\kappa}\vert\phi}\\
        F_{\vert\vec{\kappa}\vert\theta}&F_{\theta\theta}&F_{\theta\phi}\\
        F_{\vert\vec{\kappa}\vert\phi}&F_{\theta\phi}&F_{\phi\phi}
    \end{pmatrix}\\&=\sum_{X=A,B}\int d^3\rho \frac{1}{P_\nu\left(X;\vec{ \rho}|\vec{\kappa}\right)}\\&\times\begin{pmatrix}
        \frac{\partial P_\nu\left(X;\vec{ \rho}|\vec{\kappa}\right)}{\partial \vert\vec{\kappa}\vert}\\
        \frac{\partial P_\nu\left(X;\vec{ \rho}|\vec{\kappa}\right)}{\partial \theta}\\
        \frac{\partial P_\nu\left(X;\vec{ \rho}|\vec{\kappa}\right)}{\partial \phi}
    \end{pmatrix}\begin{pmatrix}
        \frac{\partial P_\nu\left(X;\vec{ \rho}|\vec{\kappa}\right)}{\partial \vert\vec{\kappa}\vert}&
        \frac{\partial P_\nu\left(X;\vec{ \rho}|\vec{\kappa}\right)}{\partial \theta}&
        \frac{\partial P_\nu\left(X;\vec{ \rho}|\vec{\kappa}\right)}{\partial \phi}
    \end{pmatrix}
    \\&=\frac{\gamma^2}{(2\pi)^{3/2}}\int d^3 \rho\mathrm{e}^{-\vert\vec{ \rho}\vert^2/2}\beta(\vec{\kappa}\cdot\vec{ \rho})\\&\times\begin{pmatrix}
       \frac{(\vec{\kappa}\cdot\vec{ \rho})^2}{\vert\vec{\kappa}\vert^2}&\frac{(\vec{\kappa}\cdot\vec{ \rho})}{\vert\vec{\kappa}\vert} \left(\vec{ \rho}\cdot\frac{\partial\vec\kappa}{\partial\theta}\right)&\frac{(\vec{\kappa}\cdot\vec{ \rho})}{\vert\vec{\kappa}\vert}\left(\vec{ \rho}\cdot\frac{\partial\vec\kappa}{\partial\phi}\right)\\
       \frac{(\vec{\kappa}\cdot\vec{ \rho})}{\vert\vec{\kappa}\vert} \left(\vec{ \rho}\cdot\frac{\partial\vec\kappa}{\partial\theta}\right)&\left(\vec{ \rho}\cdot\frac{\partial\vec\kappa}{\partial\theta}\right)^2&\left(\vec{ \rho}\cdot\frac{\partial\vec\kappa}{\partial\theta}\right)\left(\vec{ \rho}\cdot\frac{\partial\vec\kappa}{\partial\phi}\right)\\
       \frac{(\vec{\kappa}\cdot\vec{ \rho})}{\vert\vec{\kappa}\vert}\left(\vec{ \rho}\cdot\frac{\partial\vec\kappa}{\partial\phi}\right)&\left(\vec{ \rho}\cdot\frac{\partial\vec\kappa}{\partial\theta}\right)\left(\vec{ \rho}\cdot\frac{\partial\vec\kappa}{\partial\phi}\right)&\left(\vec{ \rho}\cdot\frac{\partial\vec\kappa}{\partial\phi}\right)^2
    \end{pmatrix},\label{app:fidef}
    }
\end{eqnarray}
where, $\beta(\vec{\kappa}\cdot\vec{ \rho})$ is obtained using Eq.~\eqref{maineq:quantumbeats} of the main text,
\begin{equation}
    \beta_\nu(x)=\sum_{X=A,B}\frac{1}{\zeta_{X;\nu}(x)}\left(\frac{\partial \zeta_{X;\nu}(x)}{\partial x}\right)^2=\frac{\nu^2\sin^2(x)}{1-\nu^2\cos^2(x)},
\end{equation}
which is Eq.~\eqref{eq:beta} of the main text. The off-diagonal terms of the integral in Eq.~\eqref{app:fidef} are odd functions with respect to the parameters $\frac{(\vec{\kappa}\cdot\vec{ \rho})}{\vert\vec{\kappa}\vert}$, $\frac{\partial\vec\kappa}{\partial\theta}$ and $\frac{\partial\vec\kappa}{\partial\phi}$. Therefore the Fisher information matrix is diagonal
\begin{equation}
    F_\nu(\vert\vec{\kappa}\vert,\theta,\phi)=\frac{\gamma^2}{(2\pi)^{3/2}}\int d^3 \rho\mathrm{e}^{-\vert\vec{ \rho}\vert^2/2}\beta(\vec{\kappa}\cdot\vec{ \rho})\begin{pmatrix}
       \frac{(\vec{\kappa}\cdot\vec{ \rho})^2}{\vert\vec{\kappa}\vert^2}&0&0\\
       0&\left(\vec{ \rho}\cdot\frac{\partial\vec\kappa}{\partial\theta}\right)^2&0\\
       0&0&\left(\vec{ \rho}\cdot\frac{\partial\vec\kappa}{\partial\phi}\right)^2.
    \end{pmatrix}
\end{equation}

The diagonal terms, instead, are
\begin{eqnarray}
\eqalign{
    F_{\vert\vec{\kappa}\vert\vert\vec{\kappa}\vert}&=\frac{\gamma^2}{(2\pi)^{3/2}}\int d^3 \rho\mathrm{e}^{-\vert\vec{ \rho}\vert^2/2}\beta(\vec{\kappa}\cdot\vec{ \rho})
        \frac{(\vec{\kappa}\cdot\vec{ \rho})^2}{\vert\vec{\kappa}\vert^2}\\
       &=\frac{\gamma^2}{(2\pi)^{1/2}}\int d l\mathrm{e}^{- l^2/2}\beta( l\vert\vec{\kappa}\vert)
        l^2.
       }
\end{eqnarray}
\begin{eqnarray}
\eqalign{
    F_{\theta\theta}&=\frac{\gamma^2}{(\pi)^{3/2}}\int d^3 \rho\mathrm{e}^{-\vert\vec{ \rho}\vert^2/2}\beta(\vec{\kappa}\cdot\vec{ \rho})
       \left(\vec{ \rho}\cdot\frac{\partial\vec\kappa}{\partial\theta}\right)^2\\
       &=\frac{\gamma^2\vert\vec{\kappa}\vert^2}{(\pi)^{1/2}}\int d l\mathrm{e}^{- l^2/2}\beta(\vert\vec{\kappa}\vert l)
        l^2.
       }
\end{eqnarray}
\begin{eqnarray}
\eqalign{
    F_{\phi\phi}&=\frac{\gamma^2}{(\pi)^{3/2}}\int d^3 \rho\mathrm{e}^{-\vert\vec{ \rho}\vert^2/2}\beta(\vec{\kappa}\cdot\vec{ \rho})
       \left(\vec{ \rho}\cdot\frac{\partial\vec\kappa}{\partial\phi}\right)^2\\
       &=\frac{\gamma^2\vert\vec{\kappa}\vert^2\sin^2\theta}{(\pi)^{1/2}}\int d l\mathrm{e}^{- l^2/2}\beta(\vert\vec{\kappa}\vert l)
        l^2.
       }
\end{eqnarray}
As a result, the Fisher information matrix in Eq.~\eqref{app:fidef} is
\begin{eqnarray}
F_{\nu}(\vert\vec{\kappa}\vert,\theta,\phi)&=\gamma^2\int d l f_\nu( l;\vert\vec{\kappa}\vert,\theta,\phi),\label{app:eqfi}
\end{eqnarray}
where
\begin{eqnarray}
\eqalign{
    f_\nu( l;\vert\vec{\kappa}\vert,\theta,\phi)&=\mathrm{Diag}[f_\nu( l;\vert\vec{\kappa}\vert),f_\nu( l;\theta),f_\nu( l;\phi)]\\&= \frac{\mathrm{e}^{- l^2/2}\beta_\nu(\vert\vec{\kappa}\vert l)}{(2\pi)^{1/2}}\begin{pmatrix}
         l^2&0&0\\
        0&\vert\vec{\kappa}\vert^2&0\\
        0&0&\vert\vec{\kappa}\vert^2\sin^2\theta
    \end{pmatrix},\label{app:densityfi}
}
\end{eqnarray}
These last two equations are Eq.~\eqref{eq:Fi} and Eq.~\eqref{eq:densityfi} in the main text. For $\nu=1$, we obtain Eq.~\eqref{eq:finu1} of the main text,
\begin{equation}
  F_{\nu=1}(\vert\vec{\kappa}\vert,\theta,\phi)=\gamma^2\begin{pmatrix}
        1&0&0\\
        0&\vert\vec{\kappa}\vert^2&0\\
        0&0&\vert\vec{\kappa}\vert^2\sin^2\theta
    \end{pmatrix}=\gamma^2Q(\vert\vec{\kappa}\vert,\theta,\phi) .
\end{equation}

\section{Derivation of Eq~\eqref{eq:likelihood3d} that implicitly defines the maximum likelihood estimators}
\label{app:Likelihood}
In this section, a derivation of the maximum likelihood estimators for the estimation of the parameters $(\vert\vec{\kappa}\vert,\theta,\phi)$ follows. The sensing scheme presented in this work is based on resolved-sampling measures. Considering that the i-th element of the sample has a probability $P_{\nu,\gamma}\pto{X_i;\vec{ \rho}_i\vert \vec{\kappa}}  
$ to occur, we define the likelihood as
\begin{equation}
\mathcal{L}=\prod_i P_{\nu,\gamma}\pto{X_i;\vec{ \rho}_i\vert \vec{\kappa}},  
\end{equation}
and the Log-Likelihood,
\begin{equation}
\log\mathcal{L}\propto\sum_i\log\pq{1+\alpha\pto{X_i}\nu\cos\pto{\vec{ \rho}_i\cdot\vec{\kappa}}}.
\end{equation}
In order to estimate the parameters $(\vert\vec{\kappa}\vert,\theta,\phi)$, we define three estimators such that the Log-Likelihood, as a function of these estimators, is maximized. Therefore, we impose the stationariety condition, for which
\begin{eqnarray}
    \eqalign{
        \begin{cases}
            \displaystyle\frac{\partial\log\mathcal{L}}{\partial \vert\vec{\kappa}\vert}=0\\
            \\\displaystyle\frac{\partial\log\mathcal{L}}{\partial \theta}=0\\
            \\\displaystyle\frac{\partial\log\mathcal{L}}{\partial \phi}=0
        \end{cases}\Longrightarrow
                \begin{cases}
           \displaystyle\sum_i \frac{\alpha\pto{X_i}\nu\sin\pto{\vec{ \rho}_i\cdot\vec{\kappa}}}{1+\alpha\pto{X_i}\nu\cos\pto{\vec{ \rho}_i\cdot\vec{\kappa}}}\vec{ \rho}_i\cdot\vec{\kappa}=0\\
           \\
                      \displaystyle\sum_i \frac{\alpha\pto{X_i}\nu\sin\pto{\vec{ \rho}_i\cdot\vec{\kappa}}}{1+\alpha\pto{X_i}\nu\cos\pto{\vec{ \rho}_i\cdot\vec{\kappa}}}\vec{ \rho}_i\cdot\frac{\partial\vec{\kappa}}{\partial\theta}=0\\
                      \\
                      \displaystyle\sum_i \frac{\alpha\pto{X_i}\nu\sin\pto{\vec{ \rho}_i\cdot\vec{\kappa}}}{1+\alpha\pto{X_i}\nu\cos\pto{\vec{ \rho}_i\cdot\vec{\kappa}}}\vec{ \rho}_i\cdot\frac{\partial\vec{\kappa}}{\partial\phi}=0
        \end{cases}.
    }
\end{eqnarray}
Since $\vec{\kappa}$, $\frac{\partial\vec{\kappa}}{\partial\theta}$ and $\frac{\partial\vec{\kappa}}{\partial\phi}$ are three orthogonal vectors, solving this system is equivalent to solve
\begin{equation}
  \displaystyle\sum_i \frac{\alpha\pto{X_i}\nu\sin\pto{\vec{ \rho}_i\cdot\vec{\kappa}}}{1+\alpha\pto{X_i}\nu\cos\pto{\vec{ \rho}_i\cdot\vec{\kappa}}}\vec{ \rho}_i=\begin{pmatrix}
      0\\0\\0
  \end{pmatrix}  ,
\end{equation}
which is Eq.~\eqref{eq:likelihood3d} of the main text.
By numerically solving this system for $\vert\vec{\kappa}\vert,\theta,\phi$ as functions of $\{ X_i,\vec{ \rho}_i\}$, it is possible to find the maximum likelihood estimators. 
\section{Evaluation of the Fisher information in Eq.~\eqref{eq:FIstar1d}}\label{app:lastlabelipromise}

Here, we evaluate the Fisher information in Eq.~\eqref{eq:FIstar1d} corresponding to the estimation of only one parameter performing sampling measurements which resolve only the conjugate variable associated with such parameter. In order to evaluate the output probability in this case, it is necessary to "classically average" the probability distribution in Eq.~\eqref{maineq:prob} with respect to the other two position variables that the detectors are unable to resolve. If, for example, the detectors resolve only the time delays, the probability distribution is
\begin{equation}
    P_{\nu,\gamma}(X;\tau|\xi)=\int d\kappa_x d\kappa_y P_{\nu,\gamma}(X;\vec{ \rho}|\vec{\kappa})=\frac{\gamma^2}{\sqrt{\pi}}\mathrm{e}^{-\tau^2/2}\left(1+\alpha(X)\nu^{[\xi]}\cos(\xi\tau)\right),\label{app:probxi}
\end{equation}
where the photonic distinguishability parameter corresponding to the visibility of the beating is
\begin{equation}
    \nu^{[\xi]}=\nu\exp\left[-\frac{\kappa_x^2+\kappa_y^2}{2}\right].\label{app:nuxi}
\end{equation}
Here, $\kappa_x$, $\kappa_y$ and $\xi$ are the first, second and third components of the vector $\vec{\kappa}$, while $\tau$ is the third component of the vectors $\vec{ \rho}$. Due to the symmetry of Eq.~\eqref{maineq:prob} with respect to the parameters $\kappa_x,\kappa_y,\xi$, the probability distribution when only one position variable is resolved with $*=\kappa_x,\kappa_y,\xi$, is
\begin{equation}
    P_\nu(X; C^{[*]}|*)=\frac{\gamma^2}{\sqrt{\pi}}\mathrm{e}^{- {C^{[*]}}^2/2}\left(1+\alpha(X)\nu^{[*]}\cos(* C^{[*]})\right),\label{app:probstargeneral}
\end{equation}
where $ C^{[*]}$ is the conjugate variable with respect to $*$ to which the detectors are sensitive, and 
\begin{equation}
    \nu^{[*]}=\nu\exp\left[\frac{-\vert\vec{\kappa}\vert^2}{2}\right]\exp\left[\frac{*^2}{2}\right],
\end{equation}
is the photonic distinguishability.
From Eq.~\eqref{app:probstargeneral}, we can derive the Fisher information for the parameter $*=\kappa_x,\kappa_y,\xi$ in the case only $ C^{[*]}$ is measured by the detectors. We obtain
\begin{eqnarray}
\eqalign{
    F_{\nu^{[*]}}(*)&=\sum_{X=A,B}\int d C^{[*]}\frac{1}{P_\nu(X; C^{[*]}|*)}\left(\frac{\partial P_\nu(X; C^{[*]}|*)}{\partial *}\right)^2 \\
    &=\gamma^2\int \frac{d C^{[*]}}{\sqrt{2\pi}}\mathrm{e}^{- {C^{[*]}}^2/2}\beta_{\nu^{[*]}}(* C^{[*]}) {C^{[*]}}^2.
    }
\end{eqnarray}
In this case the log-likelihood is 
\begin{equation}
    \log\mathcal{L}\propto\sum_i(1+\alpha(X_i)\nu^{[*]}\cos(* C^{[*]}_i)).
\end{equation}
therefore, the estimator of $*$ can be found by solving the equation
\begin{equation}
  \displaystyle\sum_i \frac{\alpha\pto{X_i}\nu^{[*]}\sin\pto{* C^{[*]}_i}}{1+\alpha\pto{X_i}\nu^{[*]}\cos\pto{* C^{[*]}_i}} C^{[*]}_i=0.
\end{equation}

\section{Evaluation of the Fisher information in Eq.~\eqref{eq:Fiforcomponents}}\label{app:QFINEWstar}
It is possible to evaluate the Fisher information matrix for the components of the vector $\vec{\kappa}$, namely $F_\nu(\vec{\kappa})$, by using Eq.~\eqref{maineq:prob}
\begin{eqnarray}
    \eqalign{
        F_\nu(\vec{\kappa})=\frac{\gamma^2}{(2\pi)^{3/2}}\int d^3 \rho \mathrm{e}^{- \rho^2/2}\frac{\nu^2\sin^2(\vec{ \rho}\cdot\vec{\kappa})}{1-\nu^2\cos^2(\vec{ \rho}\cdot\vec{\kappa})}\vec{ \rho}\vec{ \rho}^T.
    }
\end{eqnarray}
This 3D integral can be simplified into a 1D integral by writing the vector $\vec{ \rho}$ as a linear combination of the vectors $\vec{\kappa},\frac{\partial\vec{\kappa}}{\partial\theta},\frac{\partial\vec{\kappa}}{\partial\phi}$. We obtain
\begin{eqnarray}
    \eqalign{
       \vec{ \rho}&=(\vec{ \rho}\cdot\vec{\kappa})\frac{\vec{\kappa}}{\vert\vec{\kappa}\vert^2}+\frac{1}{\vert\frac{\partial\vec{\kappa}}{\partial\theta}\vert^2}\left(\vec{ \rho}\cdot\frac{\partial\vec{\kappa}}{\partial\theta}\right)\frac{\partial\vec{\kappa}}{\partial\theta}+\frac{1}{\vert\frac{\partial\vec{\kappa}}{\partial\phi}\vert^2}\left(\vec{ \rho}\cdot\frac{\partial\vec{\kappa}}{\partial\phi}\right)\frac{\partial\vec{\kappa}}{\partial\phi} \\
       &= l\frac{\vec{\kappa}}{\vert\vec{\kappa}\vert}+ l_\theta\frac{1}{\vert\frac{\partial\vec{\kappa}}{\partial\theta}\vert}\frac{\partial\vec{\kappa}}{\partial\theta}+ l_\phi\frac{1}{\vert\frac{\partial\vec{\kappa}}{\partial\phi}\vert}\frac{\partial\vec{\kappa}}{\partial\phi}.
    }
\end{eqnarray}
This allows us to transform the coordinates of integration from $d^3 \rho$ to $d l d l_\theta d l_\phi$. Therefore, our integral reduce to
\begin{eqnarray}
    \eqalign{
          F_\nu(\vec{\kappa})&=\frac{\gamma^2}{(2\pi)^{1/2}}\int d l \mathrm{e}^{- l^2/2}\frac{\nu^2\sin^2(\vert\vec{\kappa\vert l})}{1-\nu^2\cos^2(\vert\vec{\kappa\vert l})}\\&\times\left[ l^2 \frac{\vec{\kappa}\vec{\kappa}^2}{\vert\vec{\kappa}\vert^2}+\frac{1}{\vert\frac{\partial\vec{\kappa}}{\partial\theta}\vert^2}\frac{\partial\vec{\kappa}}{\partial\theta}\frac{\partial\vec{\kappa}^T}{\partial\theta}+\frac{1}{\vert\frac{\partial\vec{\kappa}}{\partial\phi}\vert^2}\frac{\partial\vec{\kappa}}{\partial\phi}\frac{\partial\vec{\kappa}^T}{\partial\phi}\right],
    }
\end{eqnarray}
where the terms proportional to $ l l_\theta,  l l_\phi, l_\theta l_\phi$ give an integral equal to zero, since the integrands associated to them are odd functions in the integration parameters. Furthermore, since the vectors $\vec{\kappa},\frac{\partial\vec{\kappa}}{\partial\theta},\frac{\partial\vec{\kappa}}{\partial\phi}$ are mutually orthogonal, the following identity holds,
\begin{equation}
    \frac{\vec{\kappa}\vec{\kappa}^2}{\vert\vec{\kappa}\vert^2}+\frac{1}{\vert\frac{\partial\vec{\kappa}}{\partial\theta}\vert^2}\frac{\partial\vec{\kappa}}{\partial\theta}\frac{\partial\vec{\kappa}^T}{\partial\theta}+\frac{1}{\vert\frac{\partial\vec{\kappa}}{\partial\phi}\vert^2}\frac{\partial\vec{\kappa}}{\partial\phi}\frac{\partial\vec{\kappa}^T}{\partial\phi}=I
\end{equation}
which allows us to write the integrand only with two terms, one related to the identity matrix, and one related to the projector $\frac{\vec{\kappa}\vec{\kappa}^T}{\vert\vec{\kappa}\vert^2}$, i.e.,
\begin{eqnarray}
    \eqalign{
        F_\nu(\vec{\kappa})=\frac{\gamma^2}{\sqrt{2\pi}}\int d l \mathrm{e}^{- l^2/2}\beta_\nu(\vert\vec{\kappa}\vert l)\left[I+( l^2-1)\frac{\vec{\kappa}\vec{\kappa}^T}{\vert\vec{\kappa}\vert^2}\right],
    }
\end{eqnarray}
as in Eq.~\eqref{eq:Fiforcomponents}.

\section*{References}
\bibliographystyle{iopart-num}
\bibliography{main}

@misc{cramer1999mathematical,
  title={Mathematical methods of statistics},
  author={Cram{\'e}r, Harald},
  volume={26},
  year={1999},
  publisher={Princeton university press},
  address={Princeton, NJ}
}

@misc{rohatgi2015introduction,
  title={An introduction to probability and statistics},
  author={Rohatgi, Vijay K and Saleh, AK Md Ehsanes},
  year={2015},
  publisher={John Wiley \& Sons}
}

@article{helstrom1969quantum,
  title={Quantum detection and estimation theory},
  author={Helstrom, Carl W},
  journal={Journal of Statistical Physics},
  volume={1},
  pages={231--252},
  year={1969},
  publisher={Springer}
}

@misc{holevo2011probabilistic,
  title={Probabilistic and statistical aspects of quantum theory},
  author={Holevo, Alexander S},
  volume={1},
  year={2011},
  publisher={Springer Science \& Business Media}
}

@article{PhysRevLett.59.2044,
  title = {Measurement of subpicosecond time intervals between two photons by interference},
  author = {Hong, C. K. and Ou, Z. Y. and Mandel, L.},
  journal = {Phys. Rev. Lett.},
  volume = {59},
  issue = {18},
  pages = {2044--2046},
  numpages = {0},
  year = {1987},
  month = {Nov},
  publisher = {American Physical Society},
  doi = {10.1103/PhysRevLett.59.2044},
  url = {https://link.aps.org/doi/10.1103/PhysRevLett.59.2044}
}

@article{kok2007linear,
  title={Linear optical quantum computing with photonic qubits},
  author={Kok, Pieter and Munro, William J and Nemoto, Kae and Ralph, Timothy C and Dowling, Jonathan P and Milburn, Gerard J},
  journal={Reviews of modern physics},
  volume={79},
  number={1},
  pages={135--174},
  year={2007},
  publisher={APS}
}

@article{tang2014measurement,
  title={Measurement-device-independent quantum key distribution over 200 km},
  author={Tang, Yan-Lin and Yin, Hua-Lei and Chen, Si-Jing and Liu, Yang and Zhang, Wei-Jun and Jiang, Xiao and Zhang, Lu and Wang, Jian and You, Li-Xing and Guan, Jian-Yu and others},
  journal={Physical review letters},
  volume={113},
  number={19},
  pages={190501},
  year={2014},
  publisher={APS}
}

@article{sangouard2011quantum,
  title={Quantum repeaters based on atomic ensembles and linear optics},
  author={Sangouard, Nicolas and Simon, Christoph and De Riedmatten, Hugues and Gisin, Nicolas},
  journal={Reviews of Modern Physics},
  volume={83},
  number={1},
  pages={33--80},
  year={2011},
  publisher={APS}
}

@article{teich2012variations,
  title={Variations on the theme of quantum optical coherence tomography: a review},
  author={Teich, Malvin Carl and Saleh, Bahaa EA and Wong, Franco NC and Shapiro, Jeffrey H},
  journal={Quantum Information Processing},
  volume={11},
  pages={903--923},
  year={2012},
  publisher={Springer}
}

@article{lyons2018attosecond,
  title={Attosecond-resolution hong-ou-mandel interferometry},
  author={Lyons, Ashley and Knee, George C and Bolduc, Eliot and Roger, Thomas and Leach, Jonathan and Gauger, Erik M and Faccio, Daniele},
  journal={Science advances},
  volume={4},
  number={5},
  pages={eaap9416},
  year={2018},
  publisher={American Association for the Advancement of Science}
}

@article{harnchaiwat2020tracking,
  title={Tracking the polarisation state of light via Hong-Ou-Mandel interferometry},
  author={Harnchaiwat, Natapon and Zhu, Feng and Westerberg, Niclas and Gauger, Erik and Leach, Jonathan},
  journal={Optics express},
  volume={28},
  number={2},
  pages={2210--2220},
  year={2020},
  publisher={Optica Publishing Group}
}

@article{triggiani2023ultimate,
  title={Ultimate quantum sensitivity in the estimation of the delay between two interfering photons through frequency-resolving sampling},
  author={Triggiani, Danilo and Psaroudis, Giorgos and Tamma, Vincenzo},
  journal={Physical Review Applied},
  volume={19},
  number={4},
  pages={044068},
  year={2023},
  publisher={APS}
}

@article{triggiani2024estimation,
  title={Estimation with ultimate quantum precision of the transverse displacement between two photons via two-photon interference sampling measurements},
  author={Triggiani, Danilo and Tamma, Vincenzo},
  journal={Physical Review Letters},
  volume={132},
  number={18},
  pages={180802},
  year={2024},
  publisher={APS}
}

@article{fabre2021parameter,
  title={Parameter estimation of time and frequency shifts with generalized Hong-Ou-Mandel interferometry},
  author={Fabre, Nicolas and Felicetti, Simone},
  journal={Physical Review A},
  volume={104},
  number={2},
  pages={022208},
  year={2021},
  publisher={APS}
}

@article{PhysRevA.91.013830,
  title = {Spectral correlation measurements at the Hong-Ou-Mandel interference dip},
  author = {Gerrits, T. and Marsili, F. and Verma, V. B. and Shalm, L. K. and Shaw, M. and Mirin, R. P. and Nam, S. W.},
  journal = {Phys. Rev. A},
  volume = {91},
  issue = {1},
  pages = {013830},
  numpages = {7},
  year = {2015},
  month = {Jan},
  publisher = {American Physical Society},
  doi = {10.1103/PhysRevA.91.013830},
  url = {https://link.aps.org/doi/10.1103/PhysRevA.91.013830}
}

@article{Jin:15,
author = {Rui-Bo Jin and Thomas Gerrits and Mikio Fujiwara and Ryota Wakabayashi and Taro Yamashita and Shigehito Miki and Hirotaka Terai and Ryosuke Shimizu and Masahiro Takeoka and Masahide Sasaki},
journal = {Opt. Express},
number = {22},
pages = {28836--28848},
publisher = {Optica Publishing Group},
title = {Spectrally resolved Hong-Ou-Mandel interference between independent photon sources},
volume = {23},
month = {Nov},
year = {2015},
url = {https://opg.optica.org/oe/abstract.cfm?URI=oe-23-22-28836},
doi = {10.1364/OE.23.028836},
}

@article{Gianani_2018,
doi = {10.1088/2040-8986/aad01a},
url = {https://dx.doi.org/10.1088/2040-8986/aad01a},
year = {2018},
month = {jul},
publisher = {IOP Publishing},
volume = {20},
number = {8},
pages = {085201},
author = {Ilaria Gianani and Emanuele Polino and Marco Sbroscia and Adil S Rab and Emanuele Roccia and Luca Mancino and Nicoló Spagnolo and Marco Barbieri and Fabio Sciarrino},
title = {Hong–Ou–Mandel control through spectral shaping},
journal = {Journal of Optics}
}

@article{barz2012demonstration,
  title={Demonstration of blind quantum computing},
  author={Barz, Stefanie and Kashefi, Elham and Broadbent, Anne and Fitzsimons, Joseph F and Zeilinger, Anton and Walther, Philip},
  journal={science},
  volume={335},
  number={6066},
  pages={303--308},
  year={2012},
  publisher={American Association for the Advancement of Science}
}

@article{guan2015experimental,
  title={Experimental passive round-robin differential phase-shift quantum key distribution},
  author={Guan, Jian-Yu and Cao, Zhu and Liu, Yang and Shen-Tu, Guo-Liang and Pelc, Jason S and Fejer, MM and Peng, Cheng-Zhi and Ma, Xiongfeng and Zhang, Qiang and Pan, Jian-Wei},
  journal={Physical review letters},
  volume={114},
  number={18},
  pages={180502},
  year={2015},
  publisher={APS}
}

@article{hofmann2012heralded,
  title={Heralded entanglement between widely separated atoms},
  author={Hofmann, Julian and Krug, Michael and Ortegel, Norbert and G{\'e}rard, Lea and Weber, Markus and Rosenfeld, Wenjamin and Weinfurter, Harald},
  journal={Science},
  volume={337},
  number={6090},
  pages={72--75},
  year={2012},
  publisher={American Association for the Advancement of Science}
}

@inproceedings{sgobba2023optimal,
  title={Optimal Measurement of Telecom Wavelength Single Photon Polarisation via Hong-Ou-Mandel Interferometry},
  author={Sgobba, Fabrizio and Pallotti, Deborah Katia and Elefante, Arianna and Dello Russo, Stefano and Dequal, Daniele and Siciliani de Cumis, Mario and Santamaria Amato, Luigi},
  booktitle={Photonics},
  volume={10},
  number={1},
  pages={72},
  year={2023},
  organization={MDPI}
}

@article{shih1988new,
  title={New type of Einstein-Podolsky-Rosen-Bohm experiment using pairs of light quanta produced by optical parametric down conversion},
  author={Shih, YH and Alley, Carroll O},
  journal={Physical Review Letters},
  volume={61},
  number={26},
  pages={2921},
  year={1988},
  publisher={APS}
}

@article{schermelleh2019super,
  title={Super-resolution microscopy demystified},
  author={Schermelleh, Lothar and Ferrand, Alexia and Huser, Thomas and Eggeling, Christian and Sauer, Markus and Biehlmaier, Oliver and Drummen, Gregor PC},
  journal={Nature cell biology},
  volume={21},
  number={1},
  pages={72--84},
  year={2019},
  publisher={Nature Publishing Group}
}

@article{muratore2024superresolution,
  title = {Superresolution imaging of two incoherent sources via two-photon-interference sampling measurements of the transverse momenta},
  author = {Muratore, Salvatore and Triggiani, Danilo and Tamma, Vincenzo},
  journal = {Phys. Rev. Appl.},
  volume = {23},
  issue = {5},
  pages = {054033},
  numpages = {10},
  year = {2025},
  month = {May},
  publisher = {American Physical Society},
  doi = {10.1103/PhysRevApplied.23.054033},
  url = {https://link.aps.org/doi/10.1103/PhysRevApplied.23.054033}
}

@article{maggio2025quantum1,
  title={Quantum-limited estimation of the frequency shift between two interfering photons by time sampling of their quantum beats},
  author={Maggio, Luca and Triggiani, Danilo and Facchi, Paolo and Tamma, Vincenzo},
  journal={The European Physical Journal Plus},
  volume={140},
  number={10},
  pages={954},
  year={2025},
  publisher={Springer Berlin Heidelberg}
}

@article{brooks2025quantum,
   title = {Optical Sensing Near the Quantum Limit with Enhanced Dynamic Range by Resolving the Spectra of Interfering Photons},
  author = {Brooks, Russell M. J. and Maggio, Luca and Jaeken, Thomas and Ho, Joseph and Gauger, Erik M. and Tamma, Vincenzo and Fedrizzi, Alessandro},
  journal = {Phys. Rev. Lett.},
  volume = {136},
  issue = {6},
  pages = {060803},
  numpages = {7},
  year = {2026},
  month = {Feb},
  publisher = {American Physical Society},
  doi = {10.1103/6xy6-c2yd},
  url = {https://link.aps.org/doi/10.1103/6xy6-c2yd}
}

@article{guo2025quantum,
  title={Quantum beat of two single photons in the transverse momentum space},
  author={Guo, Bixiang and Chen, Ziye and Maggio, Luca and Wu, Wenbo and Liu, Shiting and Tamma, Vincenzo and Fan, Jingyun},
  journal={Physical Review A},
  volume={112},
  number={1},
  pages={013719},
  year={2025},
  publisher={APS}
}

@article{chen2023spectrally,
  title={Spectrally resolved hong-ou-mandel interferometry with discrete color entanglement},
  author={Chen, Congzhen and Chen, Yuanyuan and Chen, Lixiang},
  journal={Physical Review Applied},
  volume={19},
  number={5},
  pages={054092},
  year={2023},
  publisher={APS}
}

@article{Liu_2020,
doi = {10.1088/1751-8121/ab5d4d},
url = {https://doi.org/10.1088/1751-8121/ab5d4d},
year = {2019},
month = {dec},
publisher = {IOP Publishing},
volume = {53},
number = {2},
pages = {023001},
author = {Liu, Jing and Yuan, Haidong and Lu, Xiao-Ming and Wang, Xiaoguang},
title = {Quantum Fisher information matrix and multiparameter estimation},
journal = {Journal of Physics A: Mathematical and Theoretical}
}

@article{wang2025ultrasensitive,
    title = {Ultrasensitive transverse deflection measurement with two-photon interference},
  author = {Wang, Chaojie and Zhang, Yuning and Chen, Yuanyuan and Chen, Lixiang},
  journal = {Phys. Rev. Appl.},
  volume = {25},
  issue = {1},
  pages = {014009},
  numpages = {9},
  year = {2026},
  month = {Jan},
  publisher = {American Physical Society},
  doi = {10.1103/s7yr-2fk3},
  url = {https://link.aps.org/doi/10.1103/s7yr-2fk3}
}

@article{barer1957refractometry,
  title={Refractometry and interferometry of living cells},
  author={Barer, Robert},
  journal={Journal of the Optical Society of America},
  volume={47},
  number={6},
  pages={545--556},
  year={1957},
  publisher={Optical Society of America}
}

@article{de2023satellite,
  title={Satellite-based quantum information networks: use cases, architecture, and roadmap},
  author={de Forges de Parny, Laurent and Alibart, Olivier and Debaud, Julien and Gressani, Sacha and Lagarrigue, Alek and Martin, Anthony and Metrat, Alexandre and Schiavon, Matteo and Troisi, Tess and Diamanti, Eleni and others},
  journal={Communications Physics},
  volume={6},
  number={1},
  pages={12},
  year={2023},
  publisher={Nature Publishing Group UK London}
}

@article{maggio2026ultimate,
  title={Ultimate quantum sensitivity in the 3D relative localisation of two single-photon emitters via two-photon interference},
  author={Maggio, Luca and Tamma, Vincenzo},
  journal={The European Physical Journal Plus},
  volume={141},
  number={1},
  pages={66},
  year={2026},
  publisher={Springer}
}

@article{pont2022quantifying,
  title={Quantifying n-photon indistinguishability with a cyclic integrated interferometer},
  author={Pont, Mathias and Albiero, Riccardo and Thomas, Sarah E and Spagnolo, Nicol{\`o} and Ceccarelli, Francesco and Corrielli, Giacomo and Brieussel, Alexandre and Somaschi, Niccolo and Huet, H{\^e}lio and Harouri, Abdelmounaim and others},
  journal={Physical Review X},
  volume={12},
  number={3},
  pages={031033},
  year={2022},
  publisher={APS}
}

@article{ding2025high,
  title={High-efficiency single-photon source above the loss-tolerant threshold for efficient linear optical quantum computing},
  author={Ding, Xing and Guo, Yong-Peng and Xu, Mo-Chi and Liu, Run-Ze and Zou, Geng-Yan and Zhao, Jun-Yi and Ge, Zhen-Xuan and Zhang, Qi-Hang and Liu, Hua-Liang and Wang, Lin-Jun and others},
  journal={Nature Photonics},
  pages={1--5},
  year={2025},
  publisher={Nature Publishing Group UK London}
}

@book{kato2013perturbation,
  title={Perturbation theory for linear operators},
  author={Kato, Tosio},
  volume={132},
  year={2013},
  publisher={Springer Science \& Business Media}
}

@article{tragaardh2015simple,
  title={A simple but precise method for quantitative measurement of the quality of the laser focus in a scanning optical microscope},
  author={Tr{\"a}g{\aa}rdh, J and Macrae, K and Travis, C and Amor, R and Norris, G and Wilson, SH and Oppo, G-L and McConnell, G},
  journal={Journal of Microscopy},
  volume={259},
  number={1},
  pages={66--73},
  year={2015},
  publisher={Wiley Online Library}
}

@article{yew2013temporally,
  title={Temporally focused wide-field two-photon microscopy: Paraxial to vectorial},
  author={Yew, Elijah YS and Sheppard, Colin JR and So, Peter TC},
  journal={Optics express},
  volume={21},
  number={10},
  pages={12951--12963},
  year={2013},
  publisher={Optical Society of America}
}

@article{Nemoto1990NonparaxialGB,
  title={Nonparaxial Gaussian beams.},
  author={Shojiro Nemoto},
  journal={Applied optics},
  year={1990},
  volume={29 13},
  pages={
          1940-6
        },
  url={https://api.semanticscholar.org/CorpusID:8143682}
}

@article{alda2003laser,
  title={Laser and Gaussian beam propagation and transformation},
  author={Alda, Javier},
  journal={Encyclopedia of optical engineering},
  volume={999},
  pages={1013},
  year={2003},
  publisher={New York}
}

@article{lee2016spatial,
  title={Spatial and spectral properties of entangled photons from spontaneous parametric down-conversion with a focused pump},
  author={Lee, Jong-Chan and Kim, Yoon-Ho},
  journal={Optics Communications},
  volume={366},
  pages={442--450},
  year={2016},
  publisher={Elsevier}
}

@article{patil2023anisotropic,
  title={Anisotropic spatial entanglement},
  author={Patil, Satyajeet and Prabhakar, Shashi and Biswas, Ayan and Kumar, Ashok and Singh, Ravindra P},
  journal={Physics Letters A},
  volume={457},
  pages={128583},
  year={2023},
  publisher={Elsevier}
}

@misc{plemmons1988matrix,
  title={Matrix analysis (roger a. horn and charles r. johnson)},
  author={Plemmons, Robert J},
  year={1988},
  publisher={Society for Industrial and Applied Mathematics}
}

@inbook{Grynberg_Aspect_Fabre_2010, place={Cambridge}, title={Complement 5B: One-photon wave packet}, booktitle={Introduction to Quantum Optics: From the Semi-classical Approach to Quantized Light}, publisher={Cambridge University Press}, author={Grynberg, Gilbert and Aspect, Alain and Fabre, Claude}, year={2010}, pages={398–412}}

@article{maggio2025multi,
  title={Multi-parameter estimation of the state of two interfering photonic qubits},
  author={Maggio, Luca and Triggiani, Danilo and Facchi, Paolo and Tamma, Vincenzo},
  journal={Physica Scripta},
  volume={100},
  number={3},
  pages={035106},
  year={2025},
  publisher={IOP Publishing}
}

@article{edgar2012imaging,
  title={Imaging high-dimensional spatial entanglement with a camera},
  author={Edgar, Matthew P and Tasca, Daniel S and Izdebski, Frauke and Warburton, Ryan E and Leach, Jonathan and Agnew, Megan and Buller, Gerald S and Boyd, Robert W and Padgett, Miles J},
  journal={Nature communications},
  volume={3},
  number={1},
  pages={984},
  year={2012},
  publisher={Nature Publishing Group UK London}
}

@article{freericks2023measure,
  title={How to measure the momentum of single quanta},
  author={Freericks, JK},
  journal={The European Physical Journal Special Topics},
  volume={232},
  number={20},
  pages={3285--3294},
  year={2023},
  publisher={Springer}
}

@article{zamir2002proof,
  title={A proof of the Fisher information inequality via a data processing argument},
  author={Zamir, Ram},
  journal={IEEE Transactions on Information Theory},
  volume={44},
  number={3},
  pages={1246--1250},
  year={2002},
  publisher={IEEE}
}

@article{PhysRevResearch.2.033191,
  title = {Interferometric sensing of the tilt angle of a Gaussian beam},
  author = {Walborn, S. P. and Aguilar, G. H. and Saldanha, P. L. and Davidovich, L. and Filho, R. L. de Matos},
  journal = {Phys. Rev. Res.},
  volume = {2},
  issue = {3},
  pages = {033191},
  numpages = {8},
  year = {2020},
  month = {Aug},
  publisher = {American Physical Society},
  doi = {10.1103/PhysRevResearch.2.033191},
  url = {https://link.aps.org/doi/10.1103/PhysRevResearch.2.033191}
}

@article{PhysRevApplied.14.024028,
  title = {Robust Interferometric Sensing Using Two-Photon Interference},
  author = {Aguilar, G. H. and Piera, R. S. and Saldanha, P. L. and Filho, R. L. de Matos and Walborn, S. P.},
  journal = {Phys. Rev. Appl.},
  volume = {14},
  issue = {2},
  pages = {024028},
  numpages = {10},
  year = {2020},
  month = {Aug},
  publisher = {American Physical Society},
  doi = {10.1103/PhysRevApplied.14.024028},
  url = {https://link.aps.org/doi/10.1103/PhysRevApplied.14.024028}
}

@article{Fiedler1964,
author = {Fiedler, Miroslav},
journal = {Czechoslovak Mathematical Journal},
language = {eng},
number = {1},
pages = {39-51},
publisher = {Institute of Mathematics, Academy of Sciences of the Czech Republic},
title = {Relations between the diagonal elements of two mutually inverse positive definite matrices},
url = {http://eudml.org/doc/12201},
volume = {14},
year = {1964},
}

\end{document}